\documentclass[aps,prb,reprint,showpacs,superscriptaddress,amsmath,amssymb]{revtex4-1}

\usepackage{graphicx}
\usepackage{dcolumn}
\usepackage{bm}
\usepackage{hyperref}
\usepackage[outdir=./]{epstopdf}
\usepackage{comment}
\usepackage{placeins}
\usepackage{amsmath,amssymb,amsthm,mathrsfs,amsfonts,dsfont} 
\usepackage{colortbl}
\usepackage{color}
\usepackage{xcolor}
\usepackage{bm}
\usepackage{fancybox}
\usepackage{upgreek}
\usepackage{braket}
\usepackage{setspace}
\usepackage{tikz}
\usepackage{array,booktabs}
\usepackage{physics}
\usepackage{stmaryrd}
\usepackage{afterpage}
\usepackage{bbm}  

\newcolumntype{L}{@{}>{\kern\tabcolsep}l<{\kern\tabcolsep}}

\renewcommand{\vec}[1]{\mathbf{#1}}
\newcommand{\idop}{\mathbbm{1}}  
\newcommand{\tper}{t_\perp}
\newcommand{\tpar}{t_\parallel}
\newcommand{\Hper}{\hat{H}_\perp}
\newcommand{\Hpar}{\hat{H}_\parallel}

\newcommand{\FD}[1]{{\color{magenta} #1}}

\newcommand{\TG}{\textcolor{orange}}

\usepackage{lineno}
\usepackage{soul}

\begin{document}

\title{Excitons in few-layer hexagonal boron nitride: Davydov splitting and surface localization}

\author{Fulvio Paleari}
\affiliation{Physics and Material Science Research Unit, University of Luxembourg, 162a avenue de la Fa\"iencerie, L-1511 Luxembourg, Luxembourg}

\author{Thomas Galvani}
\affiliation{Physics and Material Science Research Unit, University of Luxembourg, 162a avenue de la Fa\"iencerie, L-1511 Luxembourg, Luxembourg}

\author{Hakim Amara}
\affiliation{Laboratoire d'Etude des Microstructures, ONERA-CNRS, UMR104, Universit\'e Paris-Saclay, BP 72, 92322 Ch\^atillon Cedex, France}

\author{Fran\c{c}ois Ducastelle}
\affiliation{Laboratoire d'Etude des Microstructures, ONERA-CNRS, UMR104, Universit\'e Paris-Saclay, BP 72, 92322 Ch\^atillon Cedex, France}

\author{Alejandro Molina-S\'{a}nchez}
\affiliation{Physics and Material Science Research Unit, University of Luxembourg, 162a avenue de la Fa\"iencerie, L-1511 Luxembourg, Luxembourg}
\affiliation{Institute of Materials Science (ICMUV), University of Valencia, Catedr\'{a}tico Beltr\'{a}n 2, E-46980 Valencia, Spain}

\author{Ludger Wirtz}
\affiliation{Physics and Material Science Research Unit, University of Luxembourg, 162a avenue de la Fa\"iencerie, L-1511 Luxembourg, Luxembourg}

\date{\today}

\begin{abstract}
Hexagonal boron nitride (hBN) has been attracting great attention because of its strong excitonic effects. 
Taking into account few-layer systems, we investigate theoretically the effects of the number of layers on quasiparticle energies, absorption spectra, and excitonic states, placing particular focus on the Davydov splitting of the lowest bound excitons.
We describe how the inter-layer interaction as well as the variation in electronic screening as a function of layer number $N$ affects the electronic and optical properties.
Using both \textit{ab initio} simulations and a tight-binding model for an effective Hamiltonian describing the excitons, we characterize in detail the symmetry of the excitonic wavefunctions and the selection rules for their coupling to incoming light. 
We show that for $N > 2$, one can distinguish between surface excitons that are mostly localized on the outer layers and inner excitons, leading to an asymmetry in the energy separation between split excitonic states.
In particular, the bound surface excitons lie lower in energy than their inner counterparts.
Additionally, this enables us to show how the layer thickness affects the shape of the absorption spectrum.

\end{abstract}

\maketitle

\section{Introduction}

The experimental and theoretical study of the optical properties of layered materials has rapidly become a key research activity in the fields of materials science and condensed matter physics. Due to the strongly anisotropic bonding, where stacked layers mostly interact by van der Waals forces, unusual electronic and optical features were observed.\cite{Watanabe2006,Zhao2012,Xu2013,Molina2015,Zhang2015,Wu2015,Wirtz2006,Splendiani2010,Chernikov2014,Cudazzo2011,Cudazzo2016}
For example, the transition from indirect to direct band gap when going from bulk to exfoliated few-layers and monolayer, and, in general, the presence of strongly bound excitons. In particular quasi-2D, few-layer samples display much stronger excitonic effects with respect to their bulk counterparts due to reduced electronic screening in the stacking direction. Among layered materials, hexagonal boron nitride (hBN) stands out by virtue of its very high band gap ($>6$ eV),\cite{Wirtz2005,Arnaud2006,Wirtz2008,Arnaud2008,Jaffrennou2007,Watanabe2009,Museur2011,Galambosi2011,CassaboisG.2016,WirtzRubio} which makes BN-based system amenable for the fabrication of high-efficiency UV emitters.\cite{,Watanabe2004,Kubota932}

Since the experimental development is going towards the synthesis of low-defect, few-layer hBN samples,\cite{Pierret2014,Zhao2012,Xu2013} it is relevant to understand the intrinsic optical properties of multilayer hBN in more detail. Additionally, due to the relative simplicity of its lattice geometry and electronic structure, hBN turns out to be a very good model system to study and understand a variety of optical features of 2D materials.
Several hBN-related studies on these topics are already present in the literature.\cite{Cudazzo2016,Gunlycke2016,Koskelo2017}
However, most previous works have focused either on monolayer or on bulk systems, without taking into account the properties of few-layer systems, in which the removal of the symmetry along the stacking direction plays an important role. We will present a detailed study of the optical properties of few-layer hBN systems, placing particular focus on the Davydov splitting of the lowest-bound excitons.

The concept of Davydov splitting, originally developed to describe the energy levels in clusters of identical molecules,\cite{Davydov} can be applied to molecular crystals,\cite{Dawson1975} but also to layered materials consisting of identical layers stacked on top of one another. 
Indeed, Davydov splitting of phonon frequencies is observed in transition metal dicalchogenides few-layer systems.\cite{Luo2013,Staiger2015,Molina2015,Henrique2017} Considering a monolayer, we may take into account an excitonic state $S$ with degeneracy $m$. If we start adding more layers to the system, but we keep them far enough from each other as to not interact, $S$ becomes a state with degeneracy $nm$ where $n$ is the number of layers. However, if the $n$ layers are brought closer together and start interacting, the degeneracies may be lifted and we might have $n$ $m$-fold degenerate states forming a Davydov multiplet. In bulk hBN, for example, we have a Davydov pair (as the number of atoms per unit cell are equivalent to the case $n=2$) with an energy separation of $0.06$ eV and both with a large binding energy of $0.7$ eV.\cite{Wirtz2008} 
However, only one state is optically allowed and contributes to the strong excitonic peak in the absorption spectrum. 

This paper is about determining the effect of the splitting of the excitonic states on the optical properties of boron nitride multilayers. 
We combine state-of-the-art \textit{ab initio} calculations using the Bethe-Salpeter equation (BSE) and the GW approximation from many-body perturbation theory,\cite{Onida2002,Bechstedt2015} together with a tight-binding model using localized Wannier orbitals\cite{Wannier1937} with a few free parameters. 
This work will follow closely the methodology and theoretical premises of our previous work on monolayer hBN.\cite{monolayer} 

The paper is organised as follows. In Section \ref{s:gw-bse} we give a summary of the theoretical and computational details of the \textit{ab initio} calculations, while in Section \ref{s:ai_results} we present the GW-BSE results on few-layer hBN systems.  Section \ref{s:TB} is devoted to the discussion of the tight-binding excitonic model, followed by a comparison in Section \ref{s:bi} to \textit{ab initio} results, concerning the excitonic Davydov splitting of bilayer hBN. The analysis will be extended to multilayer systems in Section \ref{s:gaps}, where we present the general effects of stacking on the electronic and optical properties of BN systems. Here we show that the excitons can be localized either on the surface or on the inner layers, and we describe the optical features with the help of a linear chain model derived from the tight-binding formalism. The main text is complemented by several appendices.

\section{\textit{Ab initio}: theoretical and computational details}\label{s:gw-bse}

Our calculations employ density functional theory (DFT)\cite{dft,Martin2004} as a starting point to obtain band energies and electronic wave functions.\footnote{The DFT calculations of the electronic structures were performed with the \textsc{Quantum ESPRESSO} computational package\cite{qe}, a plane-wave code, in the local density approximation (LDA).\cite{lda} We used norm-conserving von Barth-Car pseudopotentials.} 
A first-order perturbation theory correction is then applied to the band energies by the many-body G$_0$W$_0$ approximation,\cite{Hybertsen1986} which describes how the electronic structure is affected by an electronic excitation by considering dressed quasiparticles (QPs) instead of bare electrons. 
This is crucial to obtain correct band gaps, especially in the case of low-dimensional insulating systems. For each k-point $\mathbf{k}$ and band $n$ we have $E_{n\mathbf{k}}=\epsilon_{n\mathbf{k}}+Z_{n\mathbf{k}}\bra{n\mathbf{k}}\Sigma(\epsilon_{n\mathbf{k}})-V_{\mathrm{xc}}\ket{n\mathbf{k}}$, where $E_{n\mathbf{k}}$ is the quasiparticle energy, $\epsilon_{n\mathbf{k}}$ is the bare DFT energy and $V_{\mathrm{xc}}$ is the exchange-correlation potential from DFT.  $\Sigma_{n\mathbf{k}}$ is the self-energy operator, written in Fourier space as a frequency convolution of the single-particle Green's function and the dynamically screened Coulomb interaction (where the screening is computed in the random phase approximation, RPA).\cite{Stefanucci2013} It is evaluated at the bare DFT energies, with the quasiparticle renormalization factor $Z_{nk}$ given by $[1-\bra{n\mathbf{k}}\partial \Sigma / \partial E\ket{n\mathbf{k}} |_{E=\epsilon_{n\mathbf{k}}}]^{-1}$.

Subsequently, in order to describe absorption processes and bound electron-hole states, it is necessary to abandon the single-particle picture and turn to the Bethe-Salpeter equation (BSE)\cite{Onida2002} for the electron-hole correlation function $L$. 
In the case of absorption, and in the static approximation (i.e. screening effects are instantaneous) $L$ depends only on the incoming photon frequency $\omega$ and the BSE can be formally written as:
\begin{equation}\label{eq:BSE}
L(\omega)=L_0(\omega) + L_0KL(\omega) \; ,
\end{equation}
where $L_0$ is the independent-particle correlation function and $K$ the Bethe-Salpeter kernel. $K$ only contains two terms: (i) a statically-screened direct Coulomb interaction, which is attractive and responsible for the creation of electron-hole bound states; (ii) a bare exchange Coulomb term, which is repulsive.\cite{Bechstedt2015}
Equation \eqref{eq:BSE} can be inverted and cast into an equivalent eigenvalue problem with an effective Hamiltonian in the basis of electronic transitions: $\hat{H}_{\mathrm{exc}}\Psi_{\lambda}=E_{\lambda}\Psi_{\lambda}$, with $E_\lambda$ being the excitonic binding energies. 
If we consider only the resonant transitions from a valence band $v$ to a conduction band $c$ (the Tamm-Dancoff approximation), the eigenvalue equation can be written explicitly as\cite{Martin2016}
\begin{equation}\label{eq:Hexc}
\delta_{NN^\prime}\Omega_N\Psi^N_{\lambda}+\sum_{N^\prime}\bra{N}K\ket{N^\prime}\Psi^{N^\prime}_{\lambda}=E_\lambda \Psi^N_{\lambda}.
\end{equation} 
Here $N$ / $N^\prime$ labels a transition $(vc\mathbf{k})$ / $(v^\prime c^\prime \mathbf{k^\prime})$. 
The (diagonal) first term in the Hamiltonian is given by single-particle energy differences $\Omega_N=E_{c\mathbf{k}}-E_{v\mathbf{k}}$, while the second one, containing the Bethe-Salpeter kernel, is responsible for the mixing of all available electronic transitions. Since we are considering optical absorption, we assume the incoming momentum $\mathbf{q}$ of light to be negligible, so only vertical transitions are allowed. The six-dimensional excitonic wavefunction can now be constructed in terms of Bloch states $\varphi_{n\mathbf{k}}(\mathbf{r})=\bra{\mathbf{r}}a^\dagger_{n\mathbf{k}}\ket{\mathrm{GS}}$ and excitonic weights $\Psi^{vc\mathbf{k}}_\lambda$:
\begin{equation}\label{eq:state}
\Psi_\lambda(\mathbf{r}_e,\mathbf{r}_h) = \sum_{vc\mathbf{k}}\Psi^{vc\mathbf{k}}_\lambda \varphi_{c\mathbf{k}}(\mathbf{r}_e)\varphi^*_{v\mathbf{k}}(\mathbf{r}_h),
\end{equation} 
where we have defined $\ket{\mathrm{GS}}$ as the single-particle ground state and $a^\dagger_{n\mathbf{k}}$ as the electron creation operator, while $\mathbf{r}_e$ and $\mathbf{r}_h$ are the positions of the electron and of the hole, respectively.
In order to obtain information about optical absorption, we are interested in the imaginary part of the macroscopic dielectric function $\varepsilon_M(\omega)$, which can readily be expressed in terms of the solutions of the excitonic eigenvalue problem after the long-range component of the bare Coulomb interaction has been removed from the BSE:\cite{Gatti2013}
\begin{equation}\label{eq:eps}
\varepsilon_M(\omega)=1-\lim_{\mathbf{q}\rightarrow 0}\frac{8\pi}{q^2}\sum_\lambda \frac{|\sum_N \Psi^N_\lambda \rho^N|^2}{\omega - E_\lambda+\mathrm{i}\eta},
\end{equation}
with $f_{\lambda}=|\sum_N \Psi^N_\lambda \rho^N|^2$ being the oscillator strength of exciton $\lambda$ and $\eta$ a small positive integer. 
The quantity $\rho^N$ coincides, after the limit $q\rightarrow 0$ has been taken, with the scalar product of $\bm{q}$ and the dipole matrix element in the length gauge for transition $N$.
The imaginary part $\Im[\varepsilon_M(\omega)]\equiv \varepsilon_2(\omega)$ has peaks at the energies of the excitonic states, i.e. for $\omega=E_\lambda$. 

We have used the \texttt{Yambo} code\cite{Marini2009} for the GW and BSE calculations. 
Table \ref{t:conv} summarizes the most important parameters needed to obtain converged GW $\pi$ and $\pi^*$ bands and converged (lowest-lying) excitonic peaks.
A detailed explanation of the computational details is available in Appendix \ref{app:cdetails}.

\begin{table}
\begin{center}
\begin{ruledtabular}
\begin{tabular}{ccc}   
System \ & \ k-point mesh \ & \ states summed \\
\hline
$1$L & $24\times24\times1$ & $120$ \\
$2$L & $36\times36\times1$ & $200$ \\
$3$L & $42\times42\times1$ & $200$ \\
$5$L & $48\times48\times1$ & $350$ \\
Bulk & $18\times18\times6$ & $280$ \\
\end{tabular}
\end{ruledtabular}
\caption{Size of the $k$-point mesh and number of summed states in the GW and BSE calculations of monolayer ($1$L), bilayer ($2$L), trilayer ($3$L), pentalayer ($5$L) and bulk hexagonal boron nitride. Only the highest values used between the GW and the BSE calculations are reported. The dependence of $k$-point sampling on layer number is explained in Appendix \ref{app:cdetails}.\label{t:conv}}
\end{center}
\end{table}

\begin{figure*}
\includegraphics[width=0.65\textwidth]{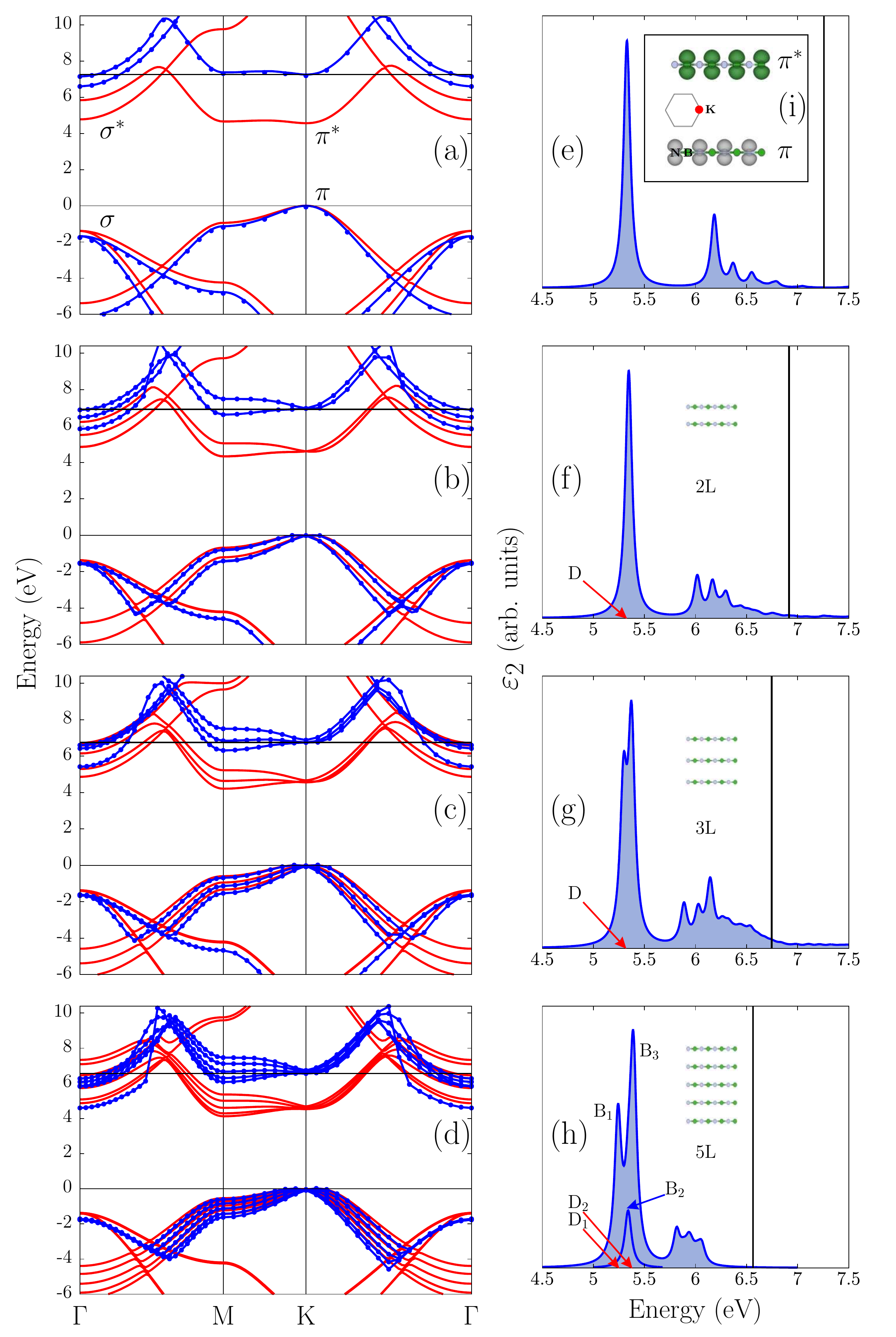}
\caption{Many-body results for monolayer [(a),(e)], bilayer [(b),(f)], trilayer [(c),(g)] and pentalayer [(d),(h)] hBN. Left panels: DFT-LDA (red) and GW (blue) band structures. In (a) the $\pi$ and $\sigma$ bands are labeled. Right panel: imaginary parts of the dielectric functions. The vertical lines represent the GW minimum vertical band gaps. The red arrows in (f), (g), (h) indicate the positions of low-energy dark (D) excitonic states. In (h), the bright peaks (B) are labeled for later comparison with Figs. \ref{fig:states} and \ref{fig:lorentzians}, and an additional bright exciton (B2), which is hidden in the main two-peak structure, is highlighted. Inset (i) shows a scheme of the hexagonal Brillouin zone, and a plot of the Bloch wavefunctions at point K for the $\pi$ and $\pi^*$ states. These states are localized on the nitrogen (gray) and boron (green) atoms, respectively. Notice that in the case of the pentalayer, the BSE was solved for states only up to the energy of $\sim 6$ eV.}
\label{fig:1}
\end{figure*}

\section{Results from \textit{ab initio} calculations}\label{s:ai_results}

The GW-BSE calculations of few-layer hBN has been done with the lattice constant fixed to the the optimized bulk value\cite{PhysRevLett.98.095503} $a=2.496$ \AA \ for all systems.
The interlayer separation was fixed at the experimental bulk value of $c=3.305$ \AA. All systems presented here are arranged in the so-called AA$^\prime$ stacking, where boron/nitrogen atoms on the various layers are vertically aligned and alternate along the stacking direction. Figure \ref{fig:1} displays the results of the GW-BSE calculations for (a)/(e) monolayer, (b)/(f) bilayer, (c)/(g) trilayer and (d)/(h) pentalayer hBN. 
In frame (a), we can see that the monolayer displays a direct band gap at k-point K. 
Except for the monolayer, all few-layer systems have indirect band gap. 
The valence $\sigma$, $\pi$ and conduction $\sigma^*$, $\pi^*$ bands are labeled. 
The electronic $\pi$ and $\pi^*$ Bloch functions at K are plotted (in gray and green, respectively) in the inset (i), to show that the valence electron remains localized on the nitrogen site (due to its larger electronegativity), while the conduction one is localized on the boron atoms. 
This means that a $\pi\rightarrow\pi^*$ electronic excitation corresponds to a hopping from nitrogen to boron (this remains mostly true while going from K to M in the Brillouin zone). As soon as a second layer is added (frame (b)), the band gap becomes indirect between a point close to K and M. 
Additionally, increased screening in the vertical direction has the effect of lowering the quasiparticle gap and the exciton binding energies with the addition of each new layer (see Fig. \ref{fig:gaps} and related discussion in Section \ref{s:gaps}). 

Around the $\Gamma$ point, a large number of parabolic conduction bands can be seen, whose energy is lower than along the MK region. 
These states are a combination of (i) $\sigma^*$ states; (ii) nearly-free electron states (NFE)\cite{Blase1995} corresponding to the bands that have $\pi^*$ character along the KM region, but that at $\Gamma$ only retain about $~30$\% $\pi^*$ character\footnote{This was confirmed by projecting these Bloch states at $\Gamma$ onto $p_z$ atomic orbitals. The overlap at K is, instead, almost $1$.}; (iii) vacuum states that slide down in energy due to the high amount of vacuum space included in the computational supercell. As the density of electronic states around $\Gamma$ increases in the case of multilayers, many (avoided) band crossings start to appear, leading to band mixing. In these cases, our G$_0$W$_0$ calculation leads to an unnatural steepness of some bands (see for example Fig. \ref{fig:1}(d)).  We believe that in order to accurately reproduce the bands in this region of the Brillouin zone a fully self-consistent GW calculation (where the electronic wave functions are also updated, instead of being kept fixed at the DFT-LDA level) should be performed.  However, these states do not participate in the optical absorption because they are either forbidden by selection rules ($\pi\rightarrow\sigma^*$, $\sigma\rightarrow\pi^*$ transitions) or the actual excitonic states are well above the band gap energy ($\sigma\rightarrow\sigma^*$), or the weight of the transitions is negligible (transitions to NFE, vacuum states). The regions in the Brillouin zone (BZ) that contribute the most to optical absorption is the one along KM, where the conduction bands are almost flat and consequently the density of states is large, although the high-energy region up to the $\pi^*$-$\sigma^*$ crossing is also relevant (more details are given in Appendix  \ref{app:transition}).

Figure \ref{fig:1} also shows the imaginary parts of the dielectric functions $\varepsilon_2(\omega)$ computed with the BSE. The vertical black lines represent the onset of the absorption continuum -- the QP band gap. 
As a result of the increased screening along the stacking direction, the binding energy reduces for increasing number of layers. 
The excitonic series in the monolayer (frame (e)) was investigated in Ref. [\onlinecite{monolayer}]. 
The shape of the excitonic wavefunction corresponding to the first peak (the lowest-bound exciton) remains the same in the multilayers. In the bilayer case, it is shown in Fig. \ref{f:2l}(b). 
If the hole is fixed above a nitrogen atom, the resulting electron density will be distributed only on the boron atoms in the same layer.

Absorption in the bilayer case, shown in frame (f), is similar to the one of the monolayer, but now a dark exciton (shown by the red arrow) appears before the main peak. These two states form a Davydov pair, which becomes a triplet in the trilayer case (frame (g)) with two bright excitons and a dark one in the middle. The pentalayer (frame (h)) shows two bright peaks as well, but a low-intensity third one (shown with a superimposed lorentzian) is hidden between them. The Davydov multiplet is completed by the presence of two dark excitons (red arrows).  

\section{Tight-binding model}\label{s:TB}

The tight-binding (TB) Hamiltonian has already been introduced in Ref.  \onlinecite{monolayer} in the case of single layer hBN. 
The DFT calculations demonstrate the localization of electrons (holes) on boron (nitrogen) sites and justifies the introduction of the tight-binding model, efficient for localized orbitals. 
Here, we give a quick overview of the model and focus on its extension to the case of multilayer hBN.
First, we consider a basis of localized $p_z$ atomic orbitals for the nitrogen ($A$) and boron ($B$) atoms: $\qty{\ket{A_\alpha, \vec{m}}, \ket{B_\beta, \vec{n}}}_{\alpha, \beta \in 1..N}$ where $\alpha, \beta$ are layer labels and $\vec{m}$ and $\vec{n}$ run over the positions of boron  and nitrogen atoms, respectively. $N$ is the number of layers. We assume the basis to be orthonormal. We now introduce an independent-particle tight-binding Hamiltonian $\hat{H}_0^{el}$, defined by: $\bra{A_\alpha, \vec{m}}\hat{H}_0^{el}\ket{A_\alpha, \vec{m}} = -\Delta$ and $\bra{B_\beta, \vec{n}}\hat{H}_0^{el}\ket{B_\beta, \vec{n}} = \Delta$, while  $\bra{A_\alpha, \vec{m}}\hat{H}_0^{el}\ket{B_\beta, \vec{n}}$ is equal to $\tpar$ if $\vec{n}$ and $\vec{m}$ are in-plane nearest neighbours, to $\tper$ if they are out-of-plane nearest neighbours, and $0$ otherwise.

From there, one can build the associated TB basis functions $\ket{A_\alpha, \vec{k}} = 1/\sqrt{M} \sum_{\vec{m} \in {\Lambda}_{h, \alpha}} e^{i \vec{k} \cdot \vec{m}} \ket{A_\alpha, \vec{m}}$ and $\ket{B_\beta, \vec{k}} = 1/\sqrt{M} \sum_{\vec{n} \in {\Lambda}_{e, \beta}} e^{i \vec{k} \cdot \vec{n}} \ket{B_\beta, \vec{n}}$,  where ${\Lambda}_{h, \alpha}$ is the physical sublattice formed by the $M$ nitrogen atoms (hole sites $h$) in layer $\alpha$, and similarly ${\Lambda}_{e, \beta}$ is the sublattice formed by boron atoms (electron sites $e$) in layer $\beta$.
This electronic TB Hamiltonian can be diagonalized to get the band structure of the $N$-layer system as a function of parameters $\Delta$, $\tpar$, $\tper$. Here, however, we are interested in the excitonic properties of the system, and we can omit this step. 
Under the assumption that electrons and holes are well localized in hBN systems, we can construct a basis of localized electron-hole excitations, and map the BSE excitonic problem of Eq. \eqref{eq:Hexc} onto a TB eigenvalue problem.

The product $\ket{\alpha, \vec{m}}_{h}\otimes\ket{\beta, \vec{n}}_{e}$ represents a specific direct-space excitation from a nitrogen atom at $\vec{m}$ in layer $\alpha$ to a boron atom at $\vec{n}$ in layer $\beta$ with electron and hole separated by a vector $\vec{R} = \vec{n}-\vec{m}$. All relevant excitonic properties can be calculated using this basis (see Appendix \ref{app:TB}).
Because of lattice translational symmetry, only the electron-hole distance vector $\vec{R} = \vec{n}-\vec{m}$ and the layers that contain the hole and of the electron, $\alpha$ and $\beta$ (since the layers are inequivalent), are of importance. 
Taking this into account, we are considering a basis of Bloch orbitals for such excitations:
\begin{equation}\label{eq:TBbasis}
\ket{\vec{R}_{\alpha, \beta}} = \frac{1}{\sqrt{M}} \sum_{\vec{m} \in {\Lambda}_{h, \alpha}} e^{i\vec{Q}\cdot\vec{m}} \ket{\alpha, \vec{m}}_{h} \otimes \ket{\beta, \vec{m}+\vec{R}}_{e},
\end{equation}
where only the $\vec{Q}=\vec{0}$ state will be considered  since we are  concerned here with direct transitions. Indirect transitions will be considered elsewhere. 
This basis consists of direct-space transitions: the state $\ket{\vec{R}_{\alpha, \beta}}$ is a Bloch orbital of all excitations with a hole in layer $\alpha$ and an electron in layer $\beta$ with an electron-hole vector $\vec{R}$.
Notice also that the elements of this basis have a geometrical interpretation: to each $\ket{\vec{R}_{\alpha, \beta}}$, one can associate a point at position $\vec{R}$ with a label $\qty(\alpha, \beta)$. 
The set of these labeled points constitutes the ``excitation lattice'' of our system, which is described in detail in Appendix \ref{app:exclat}.
From now on, the notion of first nearest neighbors (1.n.n.) will refer to points of the excitation lattice. The excitonic Hamiltonian reads: 
\begin{equation}\label{eq:TBham}
\hat{H}_X = \hat{H}_{0} + \hat{U},
\end{equation}
where $\hat{H}_0$ is the independent-particle Hamiltonian and $\hat{U}$ describes the electron-hole interaction. For the moment, we neglect the exchange interaction, so that $\hat{U}$ contains only the (screened) direct interaction. The Bethe-Salpeter Hamiltonian is (compare with Eq. \eqref{eq:Hexc}):
\begin{widetext}
\begin{equation}\label{eq:TBham2}
\bra{\vec{R}_{\alpha, \beta}}\hat{H}_X \ket{\vec{R'}_{\alpha ', \beta '}} = 
\begin{cases}
3\frac{{\tpar}^2}{\Delta} + \frac{\mathcal{B}\qty(\alpha, \beta)}{2}\frac{{\tper}^2}{\Delta} + V_{\qty(\alpha, \beta)}\qty(\vec{R}) &\mbox{if $\vec{R}_{\alpha, \beta} = \vec{R'}_{\alpha ', \beta '}$} \\
\frac{{\tpar}^2}{\Delta} \quad &\mbox{if $\vec{R}$ and $\vec{R'}$ are 1.n.n. and $\alpha=\alpha '$ and $\beta = \beta '$} \\
\frac{\tpar\tper}{\Delta} &\mbox{if $\vec{R}$ and $\vec{R'}$ are 1.n.n. and $\abs{\alpha-\alpha '}+\abs{\beta - \beta '}=1$} \\
0 &\mbox{otherwise}
\end{cases}.
\end{equation}
\end{widetext}

Here, $V_{\qty(\alpha, \beta)}\qty(\vec{R})$ is the (modified) 2D-screened Keldysh potential and $\mathcal{B}\qty(\alpha, \beta)$ is a geometrical factor. All details of the derivation of the kinetic and interaction terms of $\hat{H}_X$ are described in Appendix \ref{app:TB}.
We have shifted the energy scale by the value of the (direct) electronic gap, so that the eigenvalues of the Hamiltonian $\hat{H}_X$ are the binding energies of the excitonic states.

The excitonic states written in the basis of direct-space excitations are:
\begin{equation}
\ket{\Psi} = \sum_{\vec{R}_{\alpha, \beta}} \Psi_{\vec{R}_{\alpha, \beta}} \ket{\vec{R}_{\alpha, \beta}},
\end{equation}
obtaining the TB excitonic wavefunctions (see Eq. \eqref{eq:state} for comparison with the \textit{ab initio} expression). 
In summary, the BSE problem has been reduced to one particle moving on a lattice under the influence of an effective potential.

\section{Bilayer: \textit{Ab initio} + Tight-binding model}\label{s:bi}
\subsection{Tight-binding model}

\begin{figure}
\includegraphics[width=0.3\textwidth]{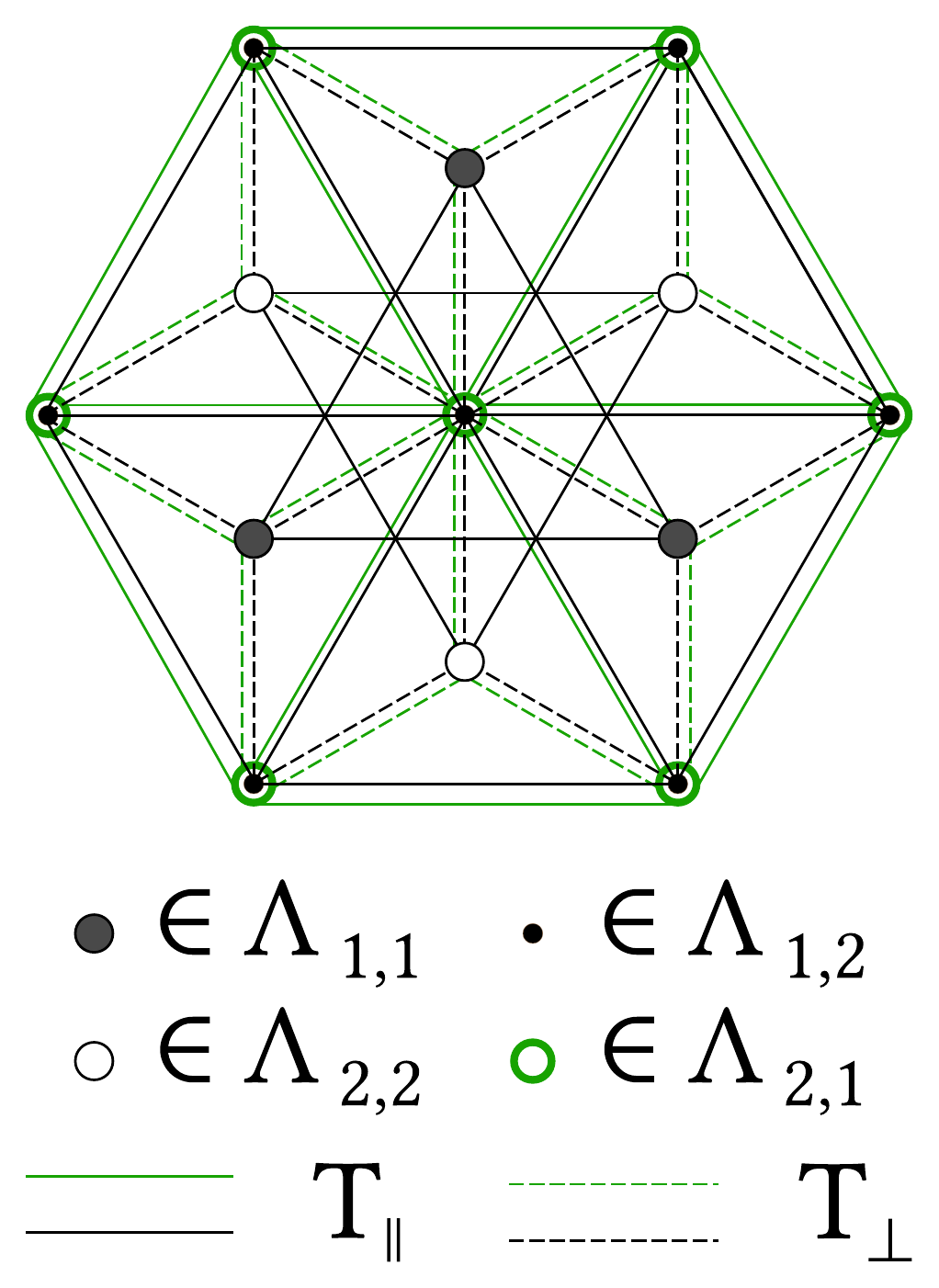}
\caption{Excitation lattice and hopping elements for the $AA^\prime$ bilayer, seen from above ($\vec{e}_z$ orthogonal to the paper plane). Dots and circles denote excitation sites, solid lines denote in-plane hoppings ($T_\parallel = \tpar^2/\Delta$) and dotted lines denote out-of-plane hoppings ($T_\perp = (\tpar \tper)/\Delta$). All sublattices are (shifted) triangular lattices and must be completed by periodicity. The lattice is composed of three planes orthogonal to $\vec{e}_z$, each containing respectively $\Lambda_{2,1}$, $\Lambda_{1,1} \cup \Lambda_{2,2}$ and $\Lambda_{1,2}$ (as defined in Eq. \eqref{eq:2Lexc}). Since $\Lambda_{2,1}$ and $\Lambda_{1,2}$ are on top of one another ($\Lambda_{1,2} = \Lambda_{2,1} + 2d\vec{e}_z$), they appear superimposed in the view from the top. For this reason, sites and hoppings connected to $\Lambda_{2,1}$ are shown in green and the corresponding hoppings slightly shifted to make them distinguishable. 
}\label{fig:2}
\end{figure}

Let us now apply the above to a detailed study of the hBN bilayer, which exhibits many of the features of the general $N$ layer case. To this end, it is useful to split the lattice of direct space excitations into $4$ different sublattices $\Lambda_{\alpha, \beta} = \qty{\vec{R}_{\alpha ', \beta '} \ | \ \qty(\alpha ', \beta ') = \qty(\alpha, \beta)}$, which results in shifted triangular lattices (see Appendix \ref{app:exclat}). 
Let $\mathcal{T}$ be the triangular lattice formed by the boron sites in layer $1$, $\vec{\tau}$ any nitrogen-boron nearest neighbour vector in layer $1$, $c$ the interlayer separation and $\vec{e}_z$ the unit vector along the stacking axis oriented from layer $1$ towards layer $2$; then:
\begin{equation}\label{eq:2Lexc}
\begin{aligned}
&\Lambda_{1, 1}=\qty{\vec{R}_{1,1} \ | \ \vec{R} \in \mathcal{T} + \vec{\tau}} \\
&\Lambda_{2, 2}=\qty{\vec{R}_{2,2} \ | \ \vec{R} \in \mathcal{T} - \vec{\tau}} \\
&\Lambda_{1, 2}=\qty{\vec{R}_{1,2} \ | \ \vec{R} \in \mathcal{T} + c \vec{e}_z} \\
&\Lambda_{2, 1}=\qty{\vec{R}_{2,1} \ | \ \vec{R} \in \mathcal{T} - c \vec{e}_z}
\end{aligned}
\end{equation}
By definition, the $\Lambda_{\alpha, \alpha}$ contain only intralayer transitions. For this reason, we call them in-plane (IP) sublattices, and excitonic states composed (mostly) of transitions from these sublattices are called intralayer or in-plane (IP) excitons. Conversely, the  $\Lambda_{\alpha, \beta}$ such that $\alpha \neq \beta$ contain only interlayer transitions which can be seen as transfering charge from one layer to another. We thus call these sublattices interlayer (IL) sublattices and the excitonic states (mostly) composed of transitions from these sublattices are denoted as interlayer (IL) excitons. In this bilayer case we have $\mathcal{B}\qty(\alpha, \beta)=2$ in the corresponding excitonic Hamiltonian of Eq. \eqref{eq:TBham2}. 
Figure \ref{fig:2} shows the structure of its hopping elements along with the structure of the lattice of excitations.
Bilayer hBN, in the $AA^\prime$ stacking, has the symmetries of the $D_{3d}$ point group, and in particular the two layers are related by inversion symmetry.
 As a consequence, $V_{\qty(1,1)}=V_{\qty(2,2)}$ and $V_{\qty(1,2)}=V_{\qty(2,1)}$. This means that the excitation lattice along with its onsite and hopping elements also possesses inversion symmetry, so that $\Lambda_{1, 1}$ and $\Lambda_{2, 2}$ are equivalent, as are $\Lambda_{1, 2}$ and $\Lambda_{2, 1}$.

Notice that neither $\Lambda_{1, 1}$ and $\Lambda_{2, 2}$ nor $\Lambda_{1, 2}$ and $\Lambda_{2, 1}$ can interact directly as no hopping element connects them. In other words, the in-plane sublattices can only interact indirectly through the interlayer sublattices and vice versa. 

\subsection{Numerical diagonalization}

We can now diagonalize $\hat{H}_X$ to obtain  the excitonic levels for the $AA'$ bilayer. All matrix elements of $\hat{H}_X$ have been specified up to the exact form of the potential $V_{\qty(\alpha, \beta)}$.  While this exact form is not required for most of our formal calculations, we require it here to perform a numerical diagonalization of $\hat{H}_X$.  As mentioned above, since our problem mostly involves two-dimensional screening, we  use a potential of the Keldysh type:\cite{Cudazzo2011,Keldysh1979}
\begin{equation*}
V_{2D}\qty(R, \rho) = \frac{\pi e^2}{2 \rho}\qty[H_0\qty(\frac{R}{\rho})-Y_0\qty(\frac{R}{\rho})] \; ,
\end{equation*}
where $\rho$ is a characteristic 2D screening length, and we set: 
\begin{align*}
V_{\qty(1, 1)}\qty(\vec{R}) &= V_{2D}\qty(R, \rho_{IP}) \\
V_{\qty(1, 2)}\qty(\vec{R}) &= V_{2D}\qty(R, \rho_{IL})  \; ,
\end{align*}
$V_{\qty(2, 2)}$ and $V_{\qty(2, 1)}$ being obtained by symmetry. 
As a result, the tight binding Hamiltonian $\hat{H}_X$ depends on four parameters: $T_\parallel = \tpar^2/\Delta$, $T_\perp = (\tpar \tper)/\Delta$, $\rho_{IP}$ and $\rho_{IL}$.

We have considered a box of $1600$ excitation sites with a cutoff of $16.05 \mbox{ \AA}$ for the hole-electron interaction. We optimize the parameters to reproduce the excitonic binding energies of the first eight \textit{ab initio} excitons (not counting degeneracies). The optimal parameters are found to be: $T_\parallel = 1.53 \mbox{ eV}$, $T_\perp = 0.454 \mbox{ eV}$, $\rho_{IP} = 12.3 \mbox{ \AA}$ and $\rho_{IL} = 16.8 \mbox{ \AA}$. The electronic gap obtained from $\hat{H}_0^{el}$ is equal to $2\Delta$, so fixing $\Delta = 3.48 \mbox{ eV}$ to reproduce the value of the \textit{ab initio} gap, we can extract the value of the corresponding electronic hoppings: $\tpar = 2.31 \mbox{ eV}$, $\tper = 0.685 \mbox{ eV}$.

\begin{table*}
\begin{center}
\begin{ruledtabular}
\begin{tabular}{ccccccccc}   
Exciton \ & \ $1 \ (\times 2)$ \ & $2 \ (\times 2)$ & $3$ & $4$ & $5 \ (\times 2)$ & $6$ & $7 \ (\times 2)$ & $8$  \\
\hline
\textit{Ab initio} & $-1.644$ & $-1.614$ & $-1.170$ & $-1.162$ & $-1.022$ & $-1.000$ & $-0.943$ & $-0.899$ \\
Tight binding & $-1.630$ & $-1.612$ & $-1.272$ & $-1.220$ & $-1.003$ & $-0.891$ & $-0.977$ & $-0.895$ \\
Bright        &   no    &  yes      &    no           &     no          &    no    &      no         &   yes     & no \\
Symmetry & $E_g$ & $E_u$  & ${A_1}_g$ & ${A_1}_u$ & $E_g$ & ${A_2}_g$ & $E_u$  & ${A_2}_u$\\
Description & $IP, 1s$ & $IP, 1s$ & $IL$ & $IL$ & $IP , 2p$ & $IL$ & $IP , 2p$ & $IL$
\end{tabular}
\end{ruledtabular}
\caption{Table of bilayer excitons. We list the \textit{ab initio} binding energies and the results of the TB fit (all the values are in eV). The optical activities and symmetries of the states are also listed, as well as their description in terms of being in-plane (IP) or interlayer (IL). For the IP excitons, a labeling of the states according to hydrogen-like energy levels is also provided.} \label{t:2L}
\end{center}
\end{table*}

By comparing the \textit{ab initio} calculations with the TB model fit, we are able to characterize the bilayer excitons in the same way it was done for the monolayer in Ref. [\onlinecite{monolayer}]. The combined \textit{ab initio}-TB results are presented in Table \ref{t:2L} up to exciton $8$ of the series. 
It can be seen that all excitons have undergone Davydov splitting into pairs of even (\textit{gerade}) and odd (\textit{ungerade}) states with respect to the inversion symmetry of the system. Explicitly, these pairs are $\qty(1,2)$, $\qty(3,8)$, $\qty(4,6)$ and $\qty(5,7)$. Pairs $\qty(1,2)$ and $\qty(5,7)$ are mostly in plane and correspond to the splitting of the first two excitons of the monolayer, respectively. The other states shown in Table \ref{t:2L} are interlayer excitons and are thus ``new" states in the sense that they are not obtained from a splitting of monolayer states. As the system possesses inversion symmetry, only  odd states can couple with light, and furthermore, for light with incoming wave vector parallel to the stacking axis (so that the field is parallel to the layers), only states with the $E$ symmetry are bright. The only optically active states are thus those of  $E_u$ symmetry. The first state is thus dark, and the main peak of the absorption spectrum comes from the second state. These selection rules are modified when incoming light is polarized along the stacking axis: in this case, only states with the $A_{2u}$ symmetry can be bright. In this case the brightest excitons are of the $IL$ type.

\subsection{Model for the Davydov splitting} 
\label{ss:mod2L}

When $\tper=0$ all sublattices decouple and $\hat{H}_X$ becomes block diagonal with respect to the sublattices. 
We denote the resulting Hamiltonian as $\Hpar$.
Let us now choose an eigenbasis $\mathcal{B}_0$ of $\Hpar$ with the following properties: all its vectors have non-vanishing intensity only on one sublattice, and the eigenvectors for the $\Lambda_{2, 2}$ and $\Lambda_{2, 1}$ blocks are the images by inversion of those of the $\Lambda_{1, 1}$ and $\Lambda_{1, 2}$ blocks, respectively.
This entails that the resulting eigenvectors are either \textit{purely} in-plane ($IP$) or \textit{purely} interlayer ($IL$) states.

We now re-introduce $\tper$ as a perturbation of $\Hpar$. Let us thus define $\Hper $ from $\Hper = \hat{H}_X - \Hpar$. 
The eigensubspaces of $\Hpar$ are in general $4$-dimensional for states that transform under the $E$ representations, and $2$-dimensional for the others. We first consider the latter. Let thus $\mathcal{E}_\Psi$ be such a two dimensional eigensubspace of $\Hpar$ corresponding to the energy $E_\Psi$: we extract from $\mathcal{B}_0$ a basis $\{\ket{\Psi_1}, \ket{\Psi_2}\}$ of $\mathcal{E}_\Psi$ such that $\ket{\Psi_1}$ and $\ket{\Psi_2}$ are images of each other by inversion and use second order degenerate perturbation theory to build an effective Hamiltonian $\hat{H}_\Psi$ in order to express the effects of the perturbation in this basis:

%
%
\begin{equation*}
\hat{H}_\Psi = E_\Psi \idop + \frac{{\tper}^2}{\Delta} \idop +
\begin{pmatrix}
g_{1,1} & g _{1,2}\\
{g _{1,2}}^* & g_{2,2}
\end{pmatrix} \; ,
\end{equation*}
where the second order terms are given by:
\begin{equation}\label{exc_splitting}
g_{i,j}=\sum_{\substack{\ket{\mu} \in \mathcal{B}_0 \\ E_\mu \neq E_\Psi}} \frac{\bra{\Psi_i}\Hper \ket{\mu}\bra{\mu}\Hper \ket{\Psi_j}}{E_\Psi - E_\mu} \; .
\end{equation}

Using inversion symmetry, it can be shown that $g_{1,2} \in \mathbb{R}$ and that $g_{1,1}=g_{2,2} \in \mathbb{R}$, so introducing the notations $g_\Psi = g_{1,2}$ and $h_\Psi = g_{1,1}=g_{2,2}$ we are left with:

\begin{equation*}
\hat{H}_\Psi = \qty(E_\Psi + \frac{{\tper}^2}{\Delta} + h_\Psi) \idop + g_\Psi
\begin{pmatrix}
0 & 1\\
1 & 0
\end{pmatrix} \; ,
\end{equation*}
from which it is clear that the states split into an even and an odd excitonic state, $\ket{\Psi_\pm}=\qty(\ket{\Psi_1} \pm \ket{\Psi_2})/\sqrt{2}$, with energies:
\begin{equation*}
E_{\Psi, \pm} = E_\Psi + \frac{{\tper}^2}{\Delta} + h_\Psi \pm g_\Psi \; ,
\end{equation*}
and this constitutes the Davydov splitting, with amplitude $2\abs{g_\Psi}$.

For states transforming under the $E$ representations, $\mathcal{E}_\Psi$ is four-dimensional: in this case we can extract from $\mathcal{B}_0$ a set of four basis states such that each sublattice contains two components transforming under the two dimensional $E$ representation. Allowing for complex wavefunctions, we can choose these components such that, on each sublattice, each of them is multiplied by $\exp (+2i\pi/3)$ or $\exp (-2i\pi/3)$ under a rotation of $2i\pi/3$. Components which transform differently under rotation cannot couple. The $4\times 4$ effective Hamiltonian can then be made block-diagonal with $2\times 2$ blocks and it can be shown that these blocks are equal. We finally recover the previous formalism. From Eq. \eqref{exc_splitting} we see that $IP$ ($IL$) states are split by interaction with $IL$ ($IP$) states respectively. Furthermore, only states of $\mathcal{B}_0$ with the same symmetry can couple. From its definition we also see that $g_\psi \propto (\tpar \tper)^2/\Delta^2$, so that the splitting scales as ${\tper}^2$. Finally we can limit the coupling to neighbouring states of  energy $E_\varphi$ so that the amplitude of the splitting $s_\Psi$ can be estimated:
\begin{equation*}
s_\Psi \sim 2 \abs{k_\Psi {\qty(\frac{\tpar \tper}{\Delta})}^2 \frac{1}{E_\Psi - E_\varphi}} \; ,
\end{equation*}
where $k_\Psi \sim \Delta^2/(\tpar \tper)^2  \bra{\Psi_i}\Hper^2 \ket{\Psi_j}$ is a dimensionless quantity. The numerical diagonalization of $\hat{H}_X$ shows that $IL$ states with  $E$ symmetry do not occur until relatively high energy into the excitonic series. On the other hand, $IL$ excitons with  $A$ symmetry occur relatively early, and so do $IP$ excitons. Assuming $k_\Psi$ to be roughly constant, this gives some qualitative understanding as to why, at least early in the excitonic series, $E$ states are  less split than $A$ states.

\subsection{Analysis of the first exciton pairs}

\begin{figure}
\includegraphics[width=0.5\textwidth]{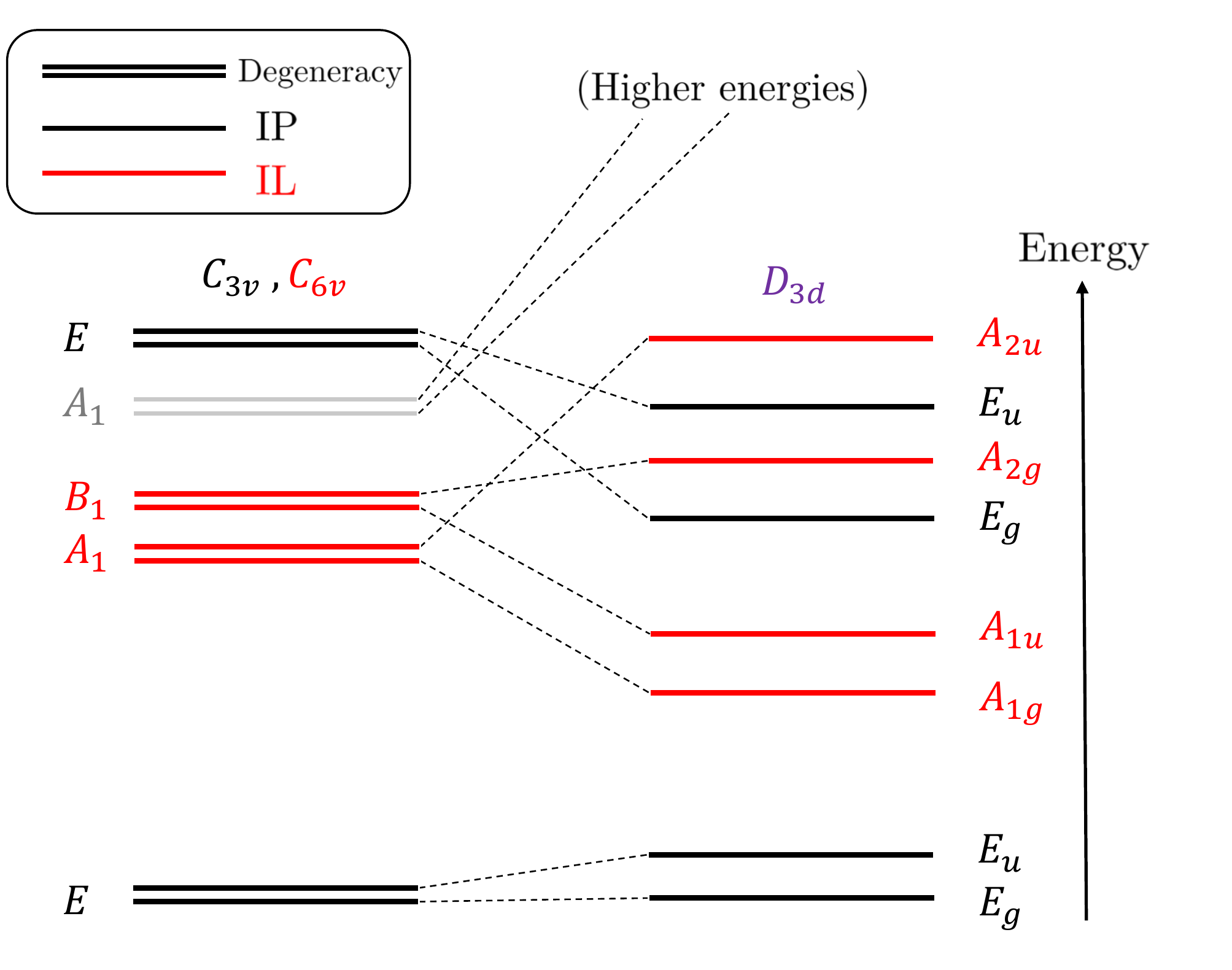}
\caption{Qualitative splitting scheme of the bilayer states presented in table \ref{t:2L}. The left side of the diagram presents eigenstates of the system without interlayer coupling ($\Hpar$), while the right side presents eigenstates of the full system. The eigenstates of $\Hpar$ were calculated using the TB model with optimal parameters but $\tper = 0$. They are labeled according to the representation of the symmetry group of the sublattice they belong to ($C_{3v}$ for the $IP$ sublattices and $C_{6v}$ for the $IL$ sublattices). The eigenstates of the full system are ordered according to their \textit{ab initio} energies and labeled according to the symmetry group of the full system, $D_{3h}$. States transforming according to two-dimensional $E$, $E_g$ or $E_u$ representations have been drawn as one-dimensional, since both components have the same behavior under splitting.}\label{f:splitting}
\end{figure}

Let us now review the eigenstates of the $AA'$ bilayer, as presented in Table \ref{t:2L}. For clarity, we discuss excitons by pairs, and separate here the states which are mostly in plane, and the mostly interlayer states. In Fig. \ref{f:splitting} we provide a scheme of the splitting of the bilayer states as obtained from the model presented above.

\subsubsection{In-plane pairs}

We start with the lowest bound pair, $\qty(1,2)$. Its excitons are of $E$ symmetry and stem from the splitting of the lowest bound monolayer states. In the monolayer, the lowest bound exciton is twice degenerate, and this is therefore also the case of excitons $1$ and $2$. Analyzing the amplitudes of the corresponding wavefunctions shows that the even (dark) state is lowest in energy, as can be seen in Fig. \ref{f:2l}(a) and (b), and we estimate $g_\Psi \approx -15 \text{ meV}  <0$ from the magnitude of the \textit{ab initio} splitting. The main peak of the bilayer absorption spectrum therefore stems from exciton $2$, which is odd and thus bright. As noted above, this pair is relatively weakly split, and with a binding energy energy of $\sim -1.6 \text{ eV}$ it is well separated from the other excitonic states of the system, which only appear about $0.4 \text{ eV}$ higher. The states of the pair are less bound than the corresponding monolayer exciton, which has a binding energy of $-1.9 \text{ eV}$:\cite{monolayer} This is likely due to the increased screening brought about by the presence of the second layer.

Similarly, the pair $\qty(5,7)$ results from the splitting of the second state of $E$ symmetry in the monolayer. There, this doubly degenerate state is responsible for the second peak in the absorption spectrum. As previously, this state splits into a pair of doubly degenerate states of symmetries $E_g$ and $E_u$ with the even state being lower in energy. The odd state, $7$, is bright and is responsible for the second peak in the absorption spectrum of the bilayer.

\subsubsection{Interlayer pairs}

It can be seen in Table \ref{t:2L} that $IL$ states are captured less accurately by the model than $IP$ states. It is possible that this difficulty originates from our use of a Keldysh-type expression to model the interlayer potential: the interlayer system is very inhomogeneous and has a finite thickness which is not negligible compared to the characteristic radii of the first few interlayer states. Nevertheless, we could recover the listed \textit{ab initio} binding energies within about $10\%$, and the qualitative agreement with the \textit{ab initio} wavefunctions is satisfying.

The lowest bound interlayer state is part of the $\qty(3, 8)$ pair.  The corresponding wavefunctions are shown in Fig. \ref{f:2l}(c) and (d). This pair exhibits a strong splitting of $271 \text{ meV}$, and again, the even state is at lowest energy. Both states are dark for incoming light polarized orthogonal to the stacking axis, but it may be noted that from the selection rules mentioned above, state $8$ would be the first peak in the absorption spectrum for light polarized parallel to the stacking axis. The $\qty(4,6)$ pair exhbits a lower, but still relatively large splitting of $161 \text{ meV}$. Analysis of the TB wavefunctions reveals that, contrary to the other pairs in table \ref{t:2L}, it is the odd state which is lower in energy. 

\subsection{\textit{Ab initio} description and comparison with TB model}
\begin{figure*}
\includegraphics[width=0.7\textwidth]{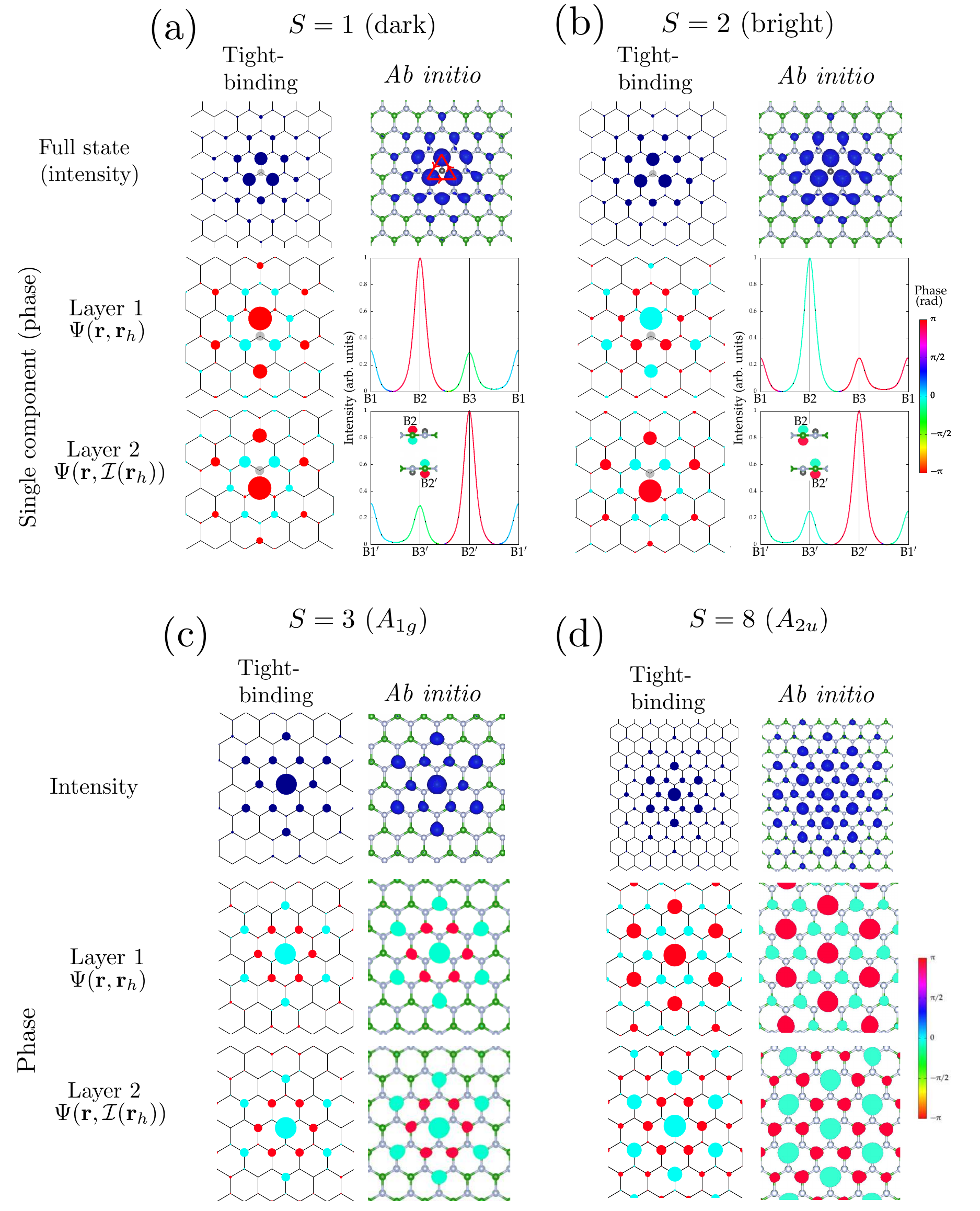}
\caption{Excitons in bilayer hBN under inversion symmetry. Tight-binding and \textit{ab initio} results. Panel (a) / (b): excitonic state $S=1$ (even, dark) / $S=2$ (odd, bright). Panel (c) / (d): excitonic state $S=3$ ($A_{1g}$) / $S=8$ ($A_{2u}$). For each panel, the tight-binding excitonic wavefunctions are shown on the left column, the \textit{ab initio} ones on the right column. In the top frames of (a) and (b) the intensities of the full doubly degenerate states on the BN lattice (hole fixed on the central nitrogen) are shown. Below, the choice of one component wavefunction in the degenerate subspace (see text and Appendix \ref{app:phase} for the procedure) permits the representation of the phase of this excitonic state. In the middle and bottom frames the phase is plotted for the electron distribution when the holes are fixed in two positions related by inversion symmetry $\mathcal{I}$ (layer 1 and layer 2). In the \textit{ab initio} case, the intensity is plotted along the triangle formed by the borons which are nearest-neighbours to the hole nitrogen (B1 and B$1^\prime=\mathcal{I}(\mathrm{B}1)$, B2 and B$2^\prime=\mathcal{I}(\mathrm{B}2)$, and B3 and B$3^\prime=\mathcal{I}(\mathrm{B}3)$: red triangle in the top frame). The phase, which rotates around the path, is shown in color scale as in the other plots. The insets show a section of the phase-intensity plots on both layers, relative to the boron with the largest intensity. In the top frames of (c) and (d) the intensities of the wavefunctions of the non-degenerate states are portrayed (since these are interlayer excitons, the hole layer is not shown as the electron density on it is very low). The phase-intensity plots are shown in the middle and bottom frames. For each exciton, two wavefunctions connected by inversion symmetry ($\mathbf{r}_{\mathrm{hole}}^\prime = \mathcal{I}(\mathbf{r}_{\mathrm{hole}}$)) are depicted, showing their respective parity [(c): even, (d): odd].}\label{f:2l}
\end{figure*}

If we look at the intensities of the lowest-bound Davydov pair, shown in the top frames of Fig. \ref{f:2l}(a) and (b), the two states appear indistinguishable (they have the shape of a lowest-bound $1s$/$E$ monolayer exciton\cite{monolayer}). 
However, state $S=1$ is optically dark, whereas state $S=2$, which lies $0.06$ eV above, is bright. This suggests that $S=1$ should be \textit{even} under inversion symmetry (dark) and $S=2$ should be \textit{odd} (bright). 
Thus, the complete symmetry analysis requires to visualize the phase of the excitonic wave function.

Since these states are doubly degenerate, we start the analysis by presenting a simpler case, the non-degenerate (dark) state $S=3$. 
Its intensity is shown in the top frames of Fig. \ref{f:2l}(c). 
This is an interlayer ($IL$) exciton: if the hole is fixed in one layer (always above a nitrogen atom, at position $r_h$), the electron density is distributed on the other (which is the only layer shown in the Figure, labeled Layer 1). 
In the middle frames of Fig. \ref{f:2l}(c) we present a phase-intensity plot of the same exciton: the values of the phase are shown in the areas with intensity greater than $5\%$. 
The phase is remarkably constant on each atom and, as expected, any two adjacent boron sites are separated by a node of the wavefunction (the phase difference is $\pi$). 
In the bottom frame of Fig. \ref{f:2l}(c) we show the same plot, but now the hole is fixed at a position $r_h^\prime=\mathcal{I}(r_h)$, where $\mathcal{I}$ is the inversion symmetry operator. The resulting electron density is now localized on the opposite layer (Layer 2) with respect to the previous case.
We can immediately see that the phase distribution does not change in the two cases: state $S=3$ is even under inversion symmetry, and we can assign it to the $A_{1g}$ representation of point group $D_{3d}$ of bilayer hBN. 
In order to find its Davydov partner, we look for an $IL$ state with the same symmetry, but odd under inversion (i.e. belonging to representation $A_{2u}$). 
We find that it is state $S=8$, represented in Fig. \ref{f:2l}(d) and listed in Table \ref{t:2L}, with a considerable Davydov splitting of $0.27$ eV. 

We are now ready to go back to the doubly-degenerate states $S=1$ and $S=2$.
In order to fully represent the phase information, it is necessary to rotate the two complex wavefunctions in the degenerate subspace until they are (almost) fully real or fully imaginary.
In the language of group theory this means that we describe the $E$ representation using a basis transforming as $x$ and $y$.
For a more detailed description of the procedure, the reader is referred to Appendix \ref{app:phase}. 

In Fig. \ref{f:2l}(a) and (b) we select one such wavefunction for each state (panel (a) for $S=1$ and (b) for $S=2$), and (in the \textit{ab initio} case) we plot a linecut of the intensity along the three boron atoms that are nearest neighbours to the nitrogen above which the hole is fixed. 
These are the sites where most of the intensity is found. 
The value of the phase (which rotates along the linecut) is shown in a color scale. 
In analogy with panels (c) and (d), the corresponding wavefunctions under inversion symmetry are plotted in the bottom frames of Fig. \ref{f:2l}(a) and (b). 
We also show sections of the phase-intensity plots for the leading peak in the insets. 
We can clearly see how $S=1$ is indeed even ($E_g$, optically forbidden) and $S=2$ is odd ($E_u$, optically active) under inversion symmetry.

\section{Trilayer and beyond: \textit{Ab initio} + Tight-binding model}\label{s:gaps}

As soon as the layer number becomes greater than two, the layers become inequivalent both in terms of degree of screening and of bonding/coordination along the stacking direction. 
This leads to various interesting features that we discuss in this Section.

\subsection{General stacking properties}

\begin{figure*}
\includegraphics[width=0.95\textwidth]{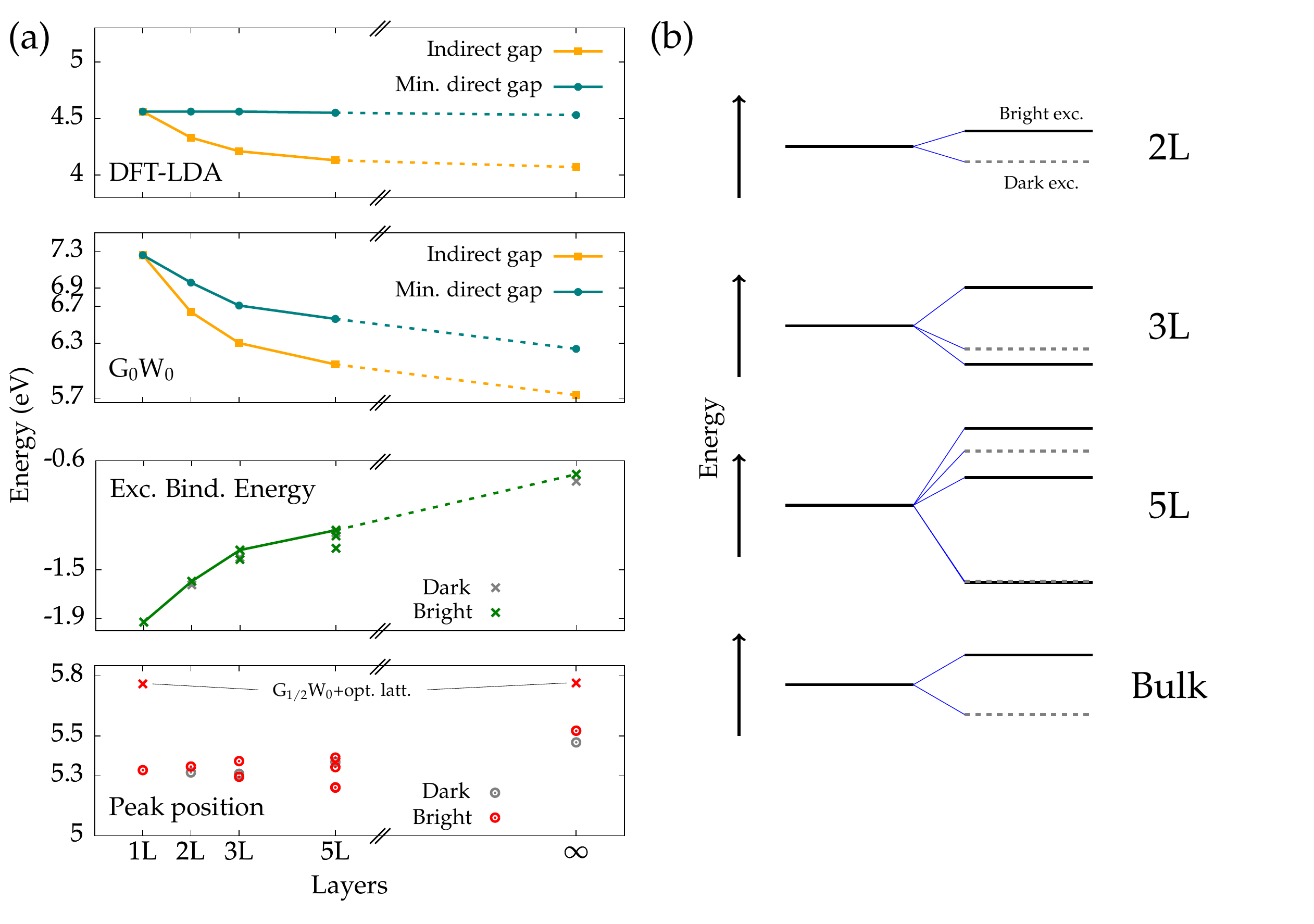}
\caption{Panel (a): DFT band gaps (top), G$_0$W$_0$ band gaps (second to top), binding energies of the lowest-bound excitons (second to bottom) and absolute excitonic peak positions (bottom) are shown as a function of the layer number. The orange squares label the band gap, the teal circles label the minimum direct gap, the green (gray) crosses / red (gray) empty circles label the binding energies / absolute peak positions of the bright (dark) excitons. The red crosses in the bottom frames represent calculations made with a semi self-consistent G$_{1/2}$W$_0$ approach and with the optimized lattice constant for the monolayer. Panel (b): scheme of the Davydov splitting in energy of the lowest-bound excitons for $N$-layer systems ($N=2,3,5,\infty$) starting from the energy of $N$ isolated monolayer excitons. Solid black (dashed gray) lines represent bright (dark) states. The energy separation within the multiplets is in scale for all systems.}
\label{fig:gaps}
\end{figure*}

The variation of the band gaps in hBN as a function of layer number is displayed in the top two frames of Fig. \ref{fig:gaps}(a).
The indirect gap (orange) and minimimum direct gap (teal) are shown both in the DFT-LDA case (upper frame) and after the GW correction (lower frame). 
In bilayer, the hybridization between the $\pi^*$ bands of the two layers, which cross at the K point slightly shifting the position of the direct band gap, has the largest effect at the M point, where the energy of the bottom band is lower than around K, giving rise to an indirect band gap. 
The energy of the bottom band at M is lowered every time the number of hybridized layers is increased, reducing the gap. 
On the other hand, the value of the direct band gap is only negligibly affected by layer stacking at the DFT level.

For both gaps the GW correction to the DFT values is huge ($\gtrsim 2$ eV). 
As the screening environment evolves from quasi-2D to 3D with layer stacking, the GW gaps decrease, converging to the bulk value. 
In particular, in the case of the minimum direct gap (relevant for optical absorption), the DFT calculation is completely unable to capture the increase in screening along the stacking direction with every added layer, giving a constant value of $4.56/4.53$ eV from monolayer to bulk. 
After the GW correction, the gap in bulk (at $6.24$ eV) is lower than the gaps in monolayer and pentalayer by $1$ and $0.3$ eV, respectively. 

The two bottom frames of Fig. \ref{fig:gaps}(a) are concerned with excitonic states. 
In the upper one, the binding energies of the lowest-bound Davydov multiplet are plotted in green (dark excitons are in gray). 
In monolayer and pentalayer, the binding energies are $1.93$ and $1.32$ eV, respectively, as opposed to $0.7$ eV in the bulk. 
By looking at the absorption spectra, we can see that the effects due to the reduction in binding energy and to the shrinking GW tend to cancel: in fact, the absolute peak positions, shown as red (bright) and gray (dark) circles in the lower frame, are almost constant, averaging around $5.3$ eV.
The position of the bulk excitons is around $5.5$ eV.

Figure \ref{fig:gaps}(b) provides for the lowest-bound exciton ($1s/E$) a scheme of the Davydov splitting from bilayer to bulk.
We make the following observations: (i) dark and bright states alternate, and (ii) in tri- and pentalayer we have a bright-dark couple at lower energy, while the rest of the multiplet lies above.
These latter states correspond to inner or ``bulk-like'' excitons (see next Section), therefore they should be compared with the bulk excitons. The bright-dark couple is made of surface excitons that have no counterpart in the bulk crystal and their relative intensity decreases to negligible values for increasing number of layers (see Appendix \ref{app:bulk}).

The bottom frame of Fig. \ref{fig:gaps}(a) shows that the energy of the bright inner peaks increases with layer number, which leads to the bulk values. However, this increasing trend might be related to the particular G$_0$W$_0$ approach. To elucidate this point, we performed simulations on monolayer and bulk using a semi self-consistent GW scheme (labeled G$_{1/2}$W$_0$), updating the band energies in $G$ during subsequent G$_0$W$_0$ runs until convergence. We obtain an additional correction to the band gap and peak positions of monolayer and bulk by $0.34$ and $0.22$ eV, respectively. We also used the LDA-optimized lattice constant for the monolayer ($2.479$ \AA) instead of the bulk one ($2.496$ \AA), which accounts for another $0.1$ eV increase in the peak energy.
The final band gap for the monolayer is $7.69$ eV, and its main excitonic peak is now almost at the same energy of the bulk one ($5.76$ eV, red crosses in Fig. \ref{fig:gaps}(a)). 
In conclusion, additional refinements in the calculations (e.g fully self-consistent GW and using the ``true'' experimental few-layer lattice constants) may lead to an inversion of the trend and show peak energies that are both higher and \textit{decreasing} towards the bulk value.

\subsection{Linear chain model}

In the following we show the relation between exciton symmetry and optical activity. The two lowest-lying states are \textit{surface} excitons (density localized  on the outer layers), while the remaining one(s) are \textit{inner} excitons (localized on the internal layers).

\subsubsection{The effective Hamiltonian}
We start with an extension to multilayer systems ($N \geq 3$) of the general TB model described in Section \ref{s:bi}.
We want to produce an effective Hamiltonian to describe the Davydov multiplets. 
It thus seems natural to proceed by defining a Hamiltonian $\Hpar$ from $\hat{H}_X$ by setting the interlayer hopping $\tper$ equal to $0$, and then build an eigenbasis of $\Hpar$ with the properties of the basis $\mathcal{B}_0$ employed in the case $N=2$.

The crucial difference with the bilayer case  stems from the fact that the layers are not all related by symmetry anymore, therefore they are no longer equivalent. 
There are two physical reasons for this: first, as can be seen in $\hat{H}_X$, transitions involving sites on the outermost ($N=1$ or $N$) layers have a lower kinetic / single-particle contribution to their on-site energy when compared to transitions involving only the inner layers. This effect is proportional to ${\tper}^2$.  As such, it is contained in $\Hper = \hat{H}_X - \Hpar$.  Secondly, transitions involving sites on the outer layers are subjected to a (gradually) lesser screening than the ones involving sites in the inner layers: the consequence of this is that the hole-electron interaction potential is more binding for transitions involving outer sites, again lowering the energy of such transitions. This effect is tied to the hole-electron interaction $\hat{U}$ and as such, is still contained in $\Hpar$.
However, it is important to note that global symmetries remain: inversion symmetry when $N$ is even and mirror symmetry when $N$ is odd.

In order to build a more symmetric basis, we define a modified Hamiltonian $\bar{H}_\parallel$ where the screening variations are averaged out (see Appendix \ref{app:1D} for more details). 
As a result, $\bar{H}_\parallel$ describes the problem of $N$ effective identical hBN layers where electrons are forbidden to hop from one layer to the other. 
It describes a system symmetric when consecutive layers are exchanged. 
We use this symmetry to build an eigenbasis $\mathcal{B}_0$ of $\bar{H}_\parallel$. 
In particular, the ground state eigensubspace of $\bar{H}_\parallel$ is spanned by $N$ copies of the  (doubly degenerate) lowest-bound monolayer exciton.
As in the $N=2$ case, we treat these copies as effectively non-degenerate states $\ket{1}$, $\ket{2}$, ..., $\ket{N}$ such that $\ket{i}$ corresponds to the effective copy on the $i^{\text{th}}$ layer, and we build an effective Hamiltonian $\hat{H}_{eff}$ in the subspace spanned by ${\qty{\ket{i}}}_{i \in \llbracket 1 ; N \rrbracket}$ to describe their splitting.
The derivation of this effective Hamiltonian is given in Appendix \ref{app:1D}. Assuming that the screening variations are only significant for the outermost layers, we obtain, up to a shift of the global energy scale:

\begin{equation}\label{eq:Heff}
\hat{H}_{eff} = -  \abs{g}\qty[ \sum_{<i,j>}\dyad{i}{j} + X \big(\dyad{1}+\dyad{N}\big)],
\end{equation}
where $\abs{g}$ describes the strength of the coupling of the states of neighbouring layers, $X$ is a dimensionless quantity characterizing the surface effect, and the sum is over nearest neighbors layers. Physically, $X$ is related to the energy difference between surface and inner layers divided by the interlayer coupling energy. This is  just a linear chain model with boundary effects.

It can be solved using standard methods. In the present case a detailed solution has been given by Puszkarski.\cite{puszkarski}  The eigenvalues are given by $E_n = -2\abs{g}\cos(k_n)$, where the $k_n$ are determined by the boundary conditions. In the case of an ideal linear chain ($X=0$), the allowed wavenumbers would be given by $k_n=n\pi/(N+1)$. Here, $X\neq 0$ \textit{a priori}, and they are determined implicitly from the relation $(\cos(k)-p(X))\sin(Nk)=r(X)\sin(k)\cos(Nk)$, with $p(X)=2X/(X^2+1)$ and $r(X)=(X^2-1)/(X^2+1)$.
It can be shown that for values of $X$ larger than a certain threshold (specifically $X\geq (N+1)/(N-1)$), this equation admits $N-2$ real solutions in $[0  \, ; \pi [$ and $2$ purely imaginary ones which correspond to surface states. One state is even, and the other is odd with respect to parity under inversion of the linear chain. As shown below, this is the crucial symmetry that controls the brightness or darkness of excitonic states in multilayer systems.

This behavior is clear in the  $X\gg 1$ regime, where we can make the  approximation that the two outer layers are completely decoupled from the $N-2$ inner layers. The approximation is relevant, since \textit{ab initio} results suggest that this might indeed be the case for $N$-layer systems ($N>3$).  The former layers will yield degenerate states with energy $-\abs{g}X$, while the latter will behave as an ideal ($X=0$) finite linear chain with $N-2$ sites with eigenenergies $E_n=-2\abs{g}\cos(n\pi/(N-1))$. Then, the coupling between outer and inner layers can be reintroduced as a perturbation. To first-order  in $1/X$ we can derive an effective Hamiltonian for the inner states in the high-energy subspace spanned by $\left\{\ket{i}\right\}_{i=2..N-1}$:

\begin{equation*}
\hat{H}_{inner} = -\abs{g}\qty[\sum_{<i, j>} \dyad{i}{j} - \frac{1}{X}\big[\dyad{1}{1}+\dyad{N}{N}\big]],
\end{equation*}
which for large $X$ is nothing more than a linear chain with weak boundary effects, that will slightly displace the energy levels and modify the states.

Let us now consider the outer surface states. Since we consider only first neighbour layer interactions, the states $\ket{1}$ and $\ket{N}$ are not coupled by second-order perturbation theory if $N>3$: we simply obtain a rigid shift of the (degenerate) energies which become equal to  $-\abs{g}(X+1/X)$. However, $\ket{1}$ and $\ket{N}$ interact indirectly via the inner states, and their splitting is seen in \textit{ab initio} calculations.
In order to describe this effect, we introduce an effective coupling integral $\gamma$  and an effective on-site energy $E_b$, so that the Hamiltonian in the $\left\{\ket{1},\ket{N}\right\}$ subspace is given by:
\begin{equation*}
\hat{H}_{outer} = -\abs{g}\big[E_b + \gamma\big(\dyad{1}{N}+\dyad{N}{1}\big) \big].
\end{equation*}

The eigenstates and eigenenergies for this two-level system are given by:
\begin{equation*}
\ket{\Psi_O, \pm} = \frac{1}{\sqrt{2}}\big( \ket{1} \pm \ket{N} \big) \quad , \quad E_{O, +} = -\abs{g}\qty(E_b \pm \gamma), 
\end{equation*}
describing a  splitting of the two surface states into an even and odd state, with a splitting width of $2\abs{g}\gamma$. As said above $E_b\simeq (X+1/X)$, and it is easily found that $\gamma \simeq 1/X^{N-2}$.

\subsubsection{Optical activity} \label{sss:ChainOptics}

The optical activity of the excitonic states  is controlled by the matrix element $\mel{\emptyset}{\hat{\mathbf{p}}}{\Psi}$ where $\mathbf{p}$ is the momentum and $\ket{\emptyset}$ and $\ket{\Psi}$ are the vacuum state and the exciton state, respectively. In general, for a state $\ket{\Psi} = \sum_{\mathbf{R}} \Psi_{\mathbf{R}} \ket{\mathbf{R}}$, we have:\cite{monolayer}

\begin{equation*}
\mel{\emptyset}{\hat{\mathbf{p}}}{\Psi} = \frac{i m_e \sqrt{M}}{\hbar} \sum_{\mathbf{R}_{\alpha, \beta}} t_{\mathbf{R}_{\alpha, \beta}} \Psi_{\mathbf{R}_{\alpha, \beta}} \mathbf{R} \; ,
\end{equation*}
where $t_{\mathbf{R}_{\alpha, \beta}} = \mel{\mathbf{m}}{\hat{H}_{0}^{el}}{\mathbf{m}+\mathbf{R}}$ for $\mathbf{m} \in \Lambda_{h, \alpha}$ such that $\mathbf{m}+\mathbf{R} \in \Lambda_{e, \beta}$ is simply the tight-binding hopping integral, i.e. $\tpar$ or $\tper$ depending on whether $\mathbf{R}$ is in plane or out of plane, respectively, or zero if $\vec{R}$ is not a boron to nitrogen nearest neighbour vector of the crystal lattice.
In the case of the first Davydov $n$-uplet the wavefunction can then be written  $\ket{\Psi} = \sum_{i=1}^N A_i \ket{i}$, so that:
\begin{equation*}
\mel{0}{\hat{\mathbf{p}}}{\Psi} =\sum_{i=1}^N A_i \mel{0}{\hat{\mathbf{p}}}{i}= \sum_{i=1}^N A_i \mathbf{d}_i,
\end{equation*}
where we have defined $\mel{0}{\hat{\mathbf{p}}}{i}\equiv \mathbf{d}_i$.

In order to proceed, we point out two important symmetries of the linear chain.
Since $\hat{H}_{eff}$ from Eq. \eqref{eq:Heff} has inversion symmetry, it follows that the components of its eigenstates are related by $A_{N-i+1}=s_{\Psi}A_i$, where $s_{\psi}$ is the parity of the state $\ket{\Psi}$ with respect to the inversion symmetry of the linear chain.
Additionally, for systems in the AA$^\prime$ stacking, the in-plane vector quantities $\mathbf{d}_i$ on each layer are related by $\mathbf{d}_{N-i+1}=-(-1)^N\mathbf{d}_i$.
We can obtain a stronger relation using the fact that by definition, $\ket{1}, \ket{2}, \ldots, \ket{N} \in\mathcal{B}_0$, and therefore $\mathbf{d}_{i+1}=-\mathbf{d}_i$ for all $i\in \llbracket 1, N-1\rrbracket$, so that letting $\vec{d}=-\vec{d}_1$, we get $\mathbf{d}_{i}={\qty(-1)}^i\mathbf{d}$ for all $i\in \llbracket 1, N\rrbracket$. We can then write:
\begin{equation*}
\mel{0}{\hat{\mathbf{p}}}{\Psi} =\frac{1-s_{\Psi}{\qty(-1)}^{N}}{2}\mathbf{d}\sum_{i=1}^N {\qty(-1)}^i A_i,
\end{equation*}
thus providing a selection rule for in-plane states: (i) if $N$ is even, the even states are dark; (ii) if $N$ is odd, the odd states are dark.
Here, the even/odd character of a state refers to its parity $s_{\Psi}$ under inversion of the chain, or, equivalently, exchange of layers $k$ and $N-k+1$ for all $k$. In physical systems this corresponds to inversion symmetry for even $N$ and to mirror symmetry with respect to the central layer for odd $N$.
The oscillator strength $f_{\Psi}$ of the bright states is then proportional to $\abs{\mel{0}{\hat{\mathbf{p}}}{\Psi}}^2 \propto \abs{S(\Psi)}^2$, with $S(\Psi)\equiv \sum_{i=1}^N (-1)^i A_{i}(\Psi)$. It can be shown that the quantities $\abs{S(\Psi)}^2$ follow the exact sum rule $\sum_{n=1}^N \abs{S\qty(\Psi_n)}^2 = {\norm{\ket{z}}}^2 = N$, where $\ket{z}$ is the vector of components $z_i ={\qty(-1)}^{i}$.

\subsection{Realistic systems}
\subsubsection{Comparison between \textit{ab initio} and the linear chain model}

\begin{figure*}
\includegraphics[width=1.0\textwidth]{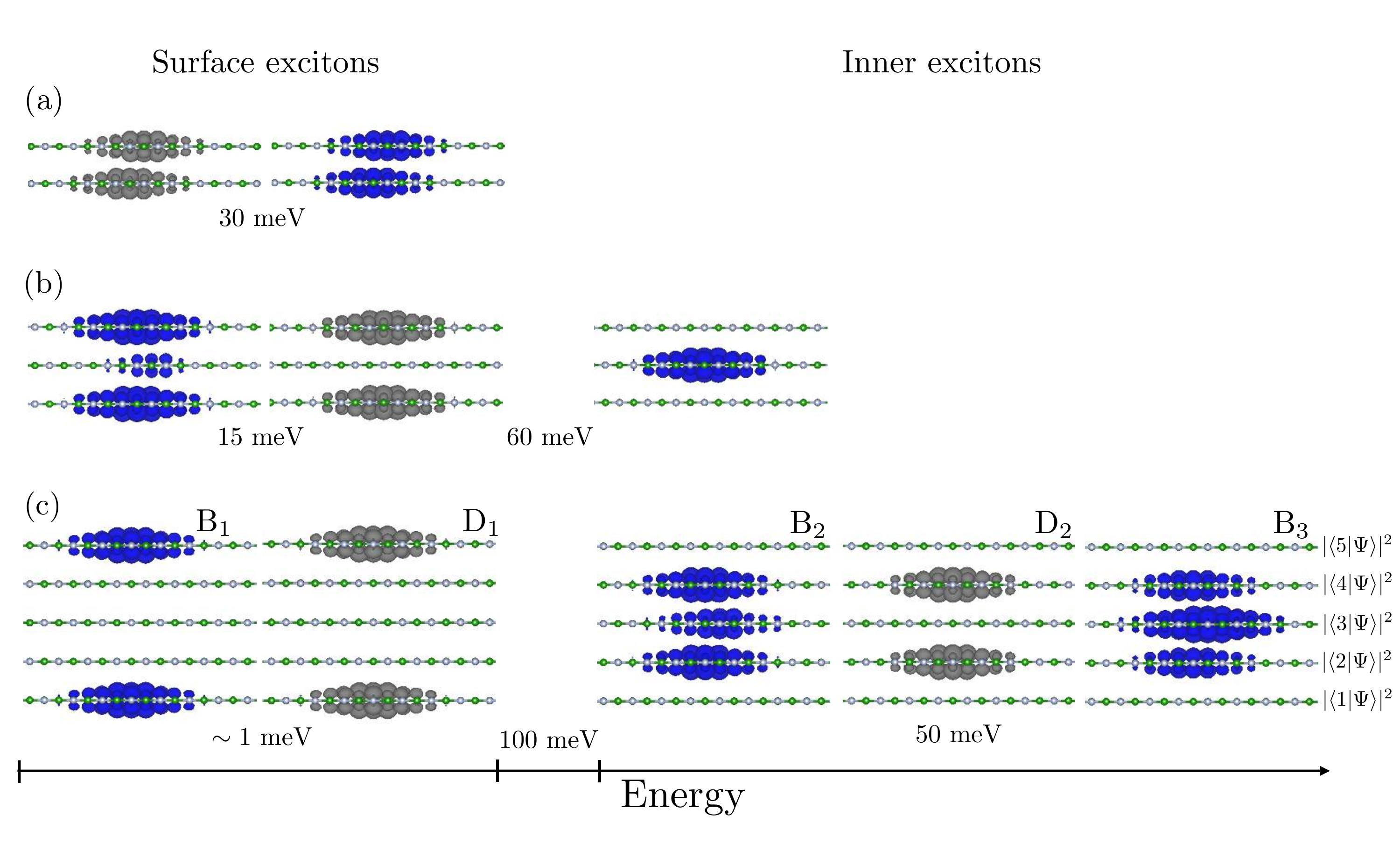}
\caption{\textit{Ab initio} simulations: lowest-bound Davydov multiplets in (a) bilayer, (b) trilayer and (c) pentalayer hBN. Each wavefunction is plotted for $N$ fixed positions of the hole ($N$ is the layer number) as explained in the text. The intensities shown in blue (gray) belong to bright (dark) excitons. The energies of the splittings between and within surface and inner exciton subsets are also given in meV. Bright (B) and dark (D) states are labeled in (c) for comparison with Figs. \ref{fig:1} and \ref{fig:lorentzians}.}\label{fig:states}
\end{figure*}

The bilayer case can be treated in the linear chain formalism, obtaining a $2\times 2$ Hamiltonian where the boundary terms only amount to a global energy shift. 
We recover the formalism of Section \ref{ss:mod2L} for the lowest bound pair.

The trilayer is the first case where boundary effects appear qualitatively in the splitting. 
The effective model, in this case, reads:
\[
\hat{H}_{eff} = -\abs{g}
\begin{pmatrix}
X & 1 & 0 \\
1 & 0 & 1 \\
0 & 1 & X
\end{pmatrix}.
\]
We only need to fit the parameter $X$ to realistic values. 
Let $E_i$ denote the energies of the three Davydov states ($i=1,2,3$). The form of $\hat{H}_{eff}$ shows that the splitting ratio $r=(E_3-E_2)/(E_2-E_1)$ depends only on $X$. \textit{Ab initio} gives the value for this ratio: $r=(60 \ \mathrm{meV})/(15 \ \mathrm{meV}) = 4$.
 By fitting $X$ on $r$ we extract a value of $X\approx 2$ (not yet in the high-$X$ regime) and we can compute the $A_i$ coefficients for each state from the TB perspective. A similar procedure can be followed for the pentalayer, and we obtain $X\approx 7.1$, which is already in the $X\gg 1$ regime.
Indeed, the structure of the absorption spectra of the hBN pentalayer in the AA$^\prime$ stacking is well described as the result of the two separated systems. First, we have two states very close in energy: these are the boundary states.  The first one is expected to be even and bright, since $N$ is odd, and the second one is odd and dark. Well above the first two states in energy ($0.1$ eV according to \textit{ab initio} results), we find a group of three states, which correspond to the inner states.  Their splitting is nearly symmetric, as expected from the model of the inner states for large $X$.  The first one of these must be even, so is bright, the second one odd thus dark, and the last one is even and thus bright. We note here that, while it is not necessary to know the exact value of $\abs{g}$ to obtain the multiplet states from the linear chain model, we can still extract its value from the amplitude of the \textit{ab initio} splittings, and we find $g_{2L} \approx -15 \text{ meV}$, $g_{3L} \approx -22 \text{ meV}$, $g_{5L} \approx -17 \text{ meV}$ and $g_{bulk} \approx -15 \text{ meV}$ respectively for the bilayer, triple-layer, pentalayer and bulk cases (see Appendix \ref{app:bulk}). This indicates that the interlayer coupling seems not to depend strongly on the number of layers.

In order to make comparisons with the \textit{ab initio} results, let us consider the excitonic wavefunction $\Psi(\mathbf{r},\mathbf{r}_h)\equiv \ket{\Psi_h}$ for a state in the lowest-bound Davydov multiplet, and the wavefunction $\ket{i}$ of the corresponding non-interacting, effective monolayer exciton localized on layer $i$. 
The subscript $h\in \llbracket 1,N\rrbracket$ denotes the fixed position of the hole in the considered layer.
We notice that, if we fix the hole in layer $j$, we have $|\Psi_j|^2 \approx |\bra{j}\ket{\Psi_h}|^2=|A_j|^2$. 
In Fig. \ref{fig:states} we show a side view of the quantity $\sum_i^N |\Psi_i|^2=\sum_i^N |A_i|^2$ (i.e. an intensity plot for $N$ different hole positions, one on each layer) for bilayer, trilayer and pentalayer hBN. 
The bright excitons are portrayed in blue, the dark ones in gray.
By comparing the figures with the tight-binding predictions we find that they are in very good qualitative agreement. 
The most important feature for multilayers is that excitons at lower (higher) energies are localized on the outer (inner) layers. 
The dark excitons are found, as expected in the case of odd layer number, to be odd with respect to the mirror symmetry of the TB linear chain (i.e. no intensity is allowed on the central layer of tri- and pentalayer).
Notice that the leading peaks in the imaginary part of the dielectric function -- see Fig. \ref{s:ai_results}(g) and (h) -- come from the excitons that are mostly localized on the central layers and are highest in energy. 

\begin{figure*}
\includegraphics[width=1.0\textwidth, trim={0cm 3cm 0cm 0cm} ]{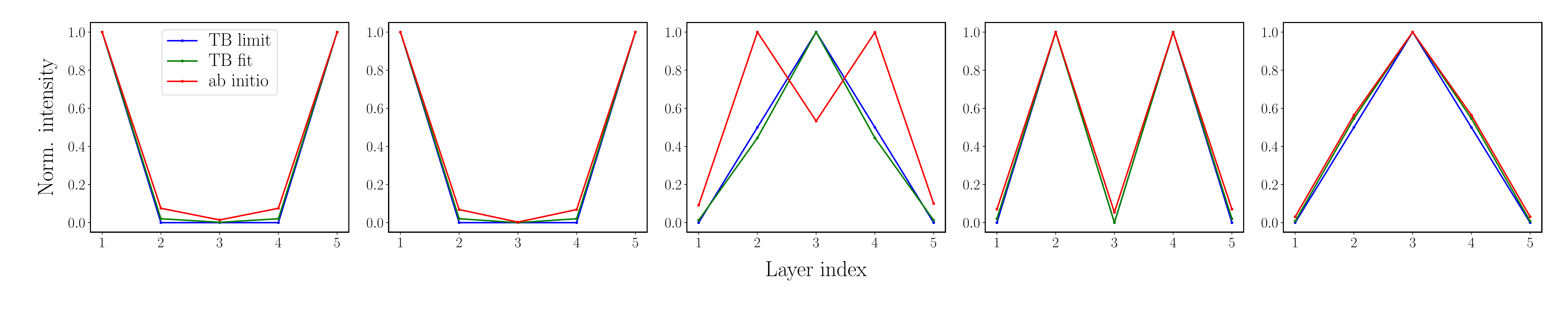}
\caption{Comparison of \textit{ab initio} (red) and tight-binding (blue, green) excitonic weights $|A_i|^2$ for each layer $i$ for the pentalayer Davydov multiplet. The blue line represents the large-$X$ limit of the linear chain model, while the green line is obtained from the diagonalization of the effective Hamiltonian Eq. \eqref{eq:Heff} via a fit of the parameters to the \textit{ab initio} binding energies.}\label{fig:5_ai_TB}
\end{figure*}

The comparison between TB and \textit{ab initio} can be made quantitative by computing the volume integrals $|\Psi_i|^2\approx |A_i|^2$ in the simulation supercell. 
The locations of the points in the numerical data grid for $|\Psi_i|^2$ must be consistent with the mirror symmetry of the linear chain, and the cell volume (grid density) must be ``converged'' to suppress numerical noise. 
The agreement is in general very good: the simple linear chain model is able to reproduce the excitonic distributions on the various layers.
Apart from two exceptions, the errors in the $A_i^{TB}$ coefficients are below $20\%$ with respect to their \textit{ab initio} counterparts. 
Larger discrepancies can appear when small, yet diffuse charge-transfer contributions for some $\Psi_i$ are present: in this case the approximate equivalence $\ket{\Psi_j}\approx A_j\ket{j}$ becomes less reliable. 
This is the case of the bright surface exciton in the trilayer (Fig. \ref{fig:states}(b)), which, according to the linear chain model, is forbidden to have a component in the central layer (i.e. $A_2^{TB}=0$ and  $A_1^{TB}=A_3^{TB}$), while in \textit{ab initio} we find $A_2^{ai}/A_1^{ai}\simeq 0.3$. 
This exciton remains nonetheless mainly localized on the surface layers.
The case of the pentalayer is summarized in Fig. \ref{fig:5_ai_TB}.
The two bright excitons localized inside the system (the third and fifth one in Figs. \ref{fig:states}(c) and \ref{fig:5_ai_TB}) are predicted to be mostly localized on the central layer, according to the linear chain model, with a $|A_3^{TB}/A_2^{TB}|^2$ ratio of $2$. 
This agrees very well with the higher-energy state (which is the brightest and thus the most important), where we find $|A_3^{ai}/A_2^{ai}|^2=1.8$, while the lower energy one has a ratio of $0.5$ and the weight distribution among the layers is inverted.
It is worth recalling that in the pentalayer the screening along the stacking direction is not constant, contrary to the tight-binding assumption of an ideal linear chain to represent the inner layers.
However, the consequent underestimation in the oscillator strength associated to this excitonic state does not influence the general agreement, as its contribution to the optical structure of the system remains very small in both the TB and \textit{ab initio} cases.

\subsubsection{Slab thickness and oscillator strength}
\begin{figure}
\includegraphics[width=0.5\textwidth]{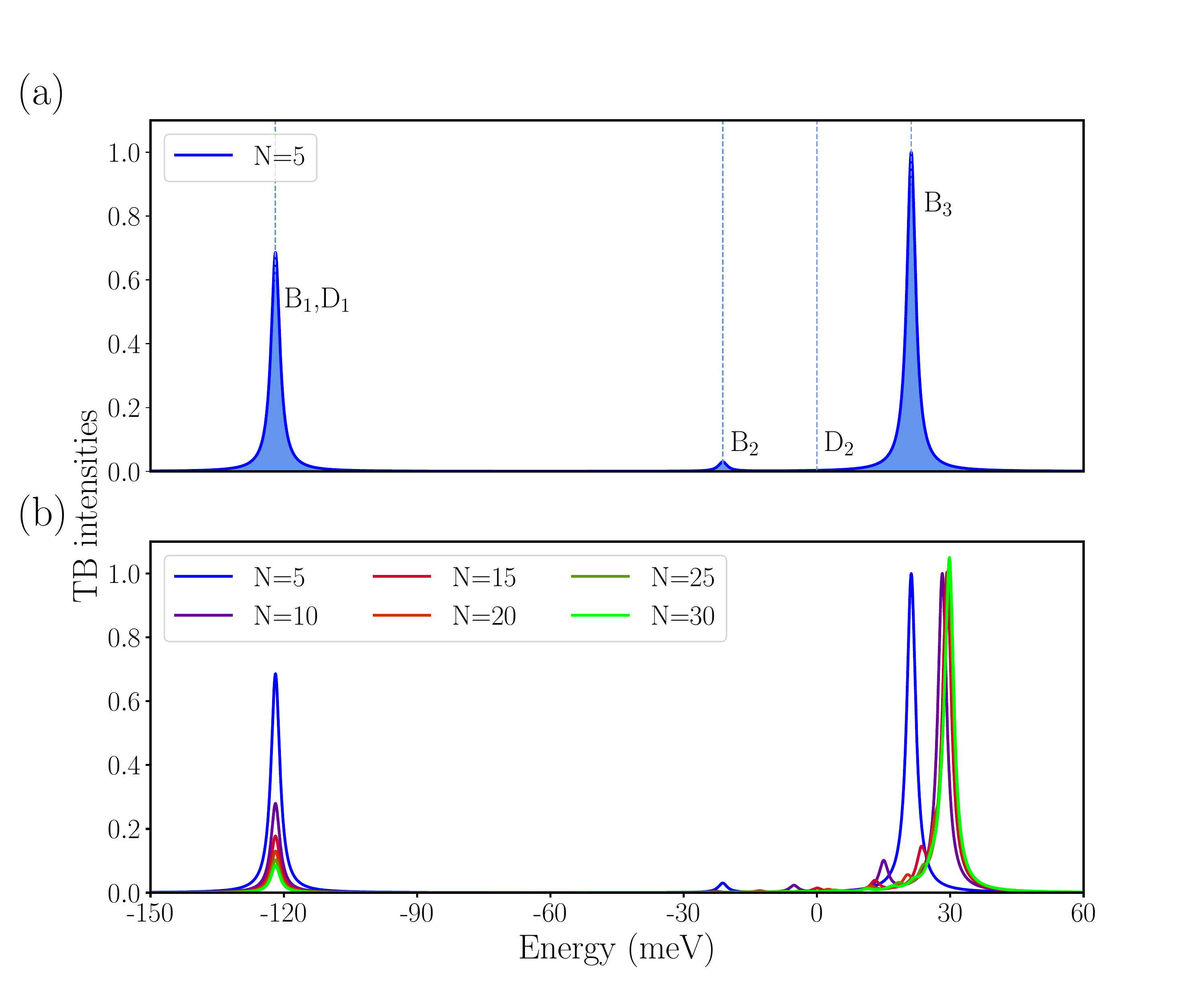}
\caption{Sum of excitonic oscillator strengths according to the linear chain model (Eq. \eqref{eq:tb_opt}) for (a) pentalayer and (b) increasing layer numbers $N$ (in different colors). The peaks are represented as lorentzians of fixed width, and $X=8 \gg 1$ is used. The energies are determined according to the cosine formula given in the text, where the global energy scale $\abs{g}$ is set to $15$ meV. The intensities are normalized with respect to the leading peak (on the right), so that the surface peak intensity (on the left) is seen to decrease with layer number. In (a), the vertical dashed lines correspond to the positions of bright (B) and dark (D) states, which are labeled for comparison with Figs. \ref{fig:1} and \ref{fig:states}.}\label{fig:lorentzians}
\end{figure}

We proceed to investigate the qualitative structure of the absorption spectrum for multilayers in the high-$X$ regime. We can then compare quantitatively the oscillator strengths $f_{\Psi} \propto \abs{S(\Psi)}^2$ from the linear chain model with the values $f_{\lambda}$ obtained from \textit{ab initio} calculations (Eq. \eqref{eq:eps}).
The outer layers will always provide a single bright peak with $\abs{S(\Psi_O)}^2 = 2$.
As for the $m=1..N-2$ inner states, which consist of an ideal linear chain, we find:
\begin{equation}\label{eq:tb_opt}
\abs{S\qty(\Psi_m)}^2 = \frac{1+{\qty(-1)}^{N+m}}{N-1} \tan[2](\frac{\pi m}{2\qty(N-1)}),
\end{equation}
so that we get an alternance of bright and dark states. 
In fact, even and odd states alternate so that if $m$ is odd, then $\ket{\Psi_m}$ is even and vice versa, and the states are either bright or dark in accordance with the selection rules presented above.
Figure \ref{fig:lorentzians} displays the resulting absorption spectrum for increasing layer number (with $X$ kept constant).
The surface peak can be seen on the left, while the inner peaks appear on the right.
The energies of the inner peaks are given by $E_m = -2\abs{g}\cos(\frac{\pi m}{N-1})$ for the values of $m$ corresponding to bright states, so that energy increases as a function of $m$. 
These peaks are concentrated in the interval $[-2\cos(\pi /(N-1))\abs{g}, 2\cos(\pi /(N-1))\abs{g}]$, and when $X$ is large enough appear as a group separated from the surface states.
For the bright states, $\abs{S\qty(\Psi_m)}^2$ is a sharply increasing function of $m$, so we see a series of increasingly bright states as we go up in energy, with the fine structure of the absorption spectrum being dominated by the last bright state, labeled $\ket{\Psi_{m^*}}$. It is always the state of highest energy in the multiplet, corresponding to $m^* = N-2$.
The position of this brightest inner peak thus tends to the upper boundary of the energy interval, $2\abs{g}$, for large values of $N$.
We can observe how the surface peaks become less prominent relative to $\ket{\Psi_{m^*}}$ as $N$ is increased.

By looking at the absorption spectrum of the pentalayer (Fig. \ref{fig:1}h), we can now identify the first peak as coming from the surface exciton, while the second one arises from the last bright inner one. 
It is clear that, as the number of layer is increased, the relative strength of the surface peak with respect to the inner one will decrease.
Therefore, the oscillator strength ratio between the two peaks provides information about the layer thickness and becomes an interesting quantity to investigate.
It is difficult to resolve experimentally each peak of this Davydov multiplet, as the energy differences involved require far-UV optical spectroscopy at very low temperatures, and may be of the same order of magnitude as other intrinsic effects (e.g. electron-electron and electron-phonon lifetimes) that give a finite width to the peaks.
However, at large $X$, the energy difference between the peak originating from the outer layers and those coming from the inner layers -- these ones appearing as a single peak without finer structure -- might be resolved experimentally.
For example, in the pentalayer, the surface-inner splitting amounts to 0.15 eV and may thus be visible in absorption spectra measured with high resolution.

In this case, we find $f_{O}/f_{m^*}=0.67$ from the 1D model in the $X\gg 1$ limit, in excellent agreement with the \textit{ab initio} result of $f_{\lambda=1}/f_{\lambda=5}=0.69$.
As mentioned at the end of the last section, the agreement is not so good for the state associated to the first inner peak, which deviates from the 1D model (see Figs. \ref{fig:1}h and \ref{fig:lorentzians}a).

Within the linear chain model, for $N\rightarrow +\infty$ and large $X$, it can be shown that the part of the absorption spectrum due to the lowest lying multiplet tends to a delta function of normalized weigth $\frac{N-2}{N}$ at energy $2\abs{g}$, and a surface peak of normalized weigth $\frac{2}{N}$. For reference, we provide a direct treatment of bulk hBN in the $AA'$ stacking in appendix \ref{app:bulk}. More precisely, for large but finite $N$, the strength of the highest energy peak $\abs{S\qty(\Psi_{m^*})}^2$ where $m^* = N-2$ is equal to:
\begin{equation*}
\abs{S\qty(\Psi_{m^*})}^2 \approx
\frac{8}{\pi^2}\qty(N-1)
\end{equation*}
so that a fraction $\frac{8}{\pi^2}\frac{N-1}{N} \sim \frac{8}{\pi^2} \sim 81 \%$ of the oscillator strength of the multiplet is due to this single state. 
Note however, that, as $N$ increases, its energy tends towards $2\abs{g}$ and more and more bright states appear arbitrarily close to this energy. These states contribute to the other $\sim 19\%$ of the oscillator strength to form the aforementioned delta function at energy $2\abs{g}$ in the limit $N\rightarrow + \infty$. 
On the other hand, the oscillator strength ratio  $f_{O}/f_{m^*}$ bright surface state over  brightest inner state decreases as $1/ (N-1)$, so that as $N$ increases towards the bulk limit, the intensity of the peak originating from the outer states gradually becomes negligible compared to that of the brightest peak from the inner states, or more generally when compared to the sum of the bright inner peaks. In fact, within the linear chain model,  $N$ can be implicitly calculated as a function of the ratio $f_{m^*}/f_{O}$ and the number of layers $N$  through:
\begin{equation}
\label{eq:ratio1Imp}
\frac{1}{N-1}\cot^2\qty(\frac{\pi}{2\qty(N-1)}) = \frac{f_{m^*}}{f_{O}}
\end{equation}
In the large $N$ limit, equation \eqref{eq:ratio1Imp} reduces to:
\begin{equation}\label{eq:ratio1}
N\approx \frac{\pi^2}{4} \frac{f_{m^*}}{f_O} +1
\end{equation}
Using \textit{ab initio} values for the pentalayer oscillator strengths, equation \ref{eq:ratio1Imp} yields $N \approx 5.0$, while its approximation, equation \ref{eq:ratio1}, yields $N\approx 4.6$.
%
%
Similar arguments provide a relationship between $N$ and the ratio $f_{in}/f_{O}$, where $f_{in} \propto \sum_{\substack{Inner \\ states}} \abs{S\qty(\Psi)}^2$ is the total oscillator strength of the inner states
\begin{equation}
\label{eq:ratio2}
N \approx 2 \qty(\frac{f_{in}}{f_O}+1).
\end{equation}
Using the \textit{ab initio} oscillator strengths in the case of the pentalayer, this formula yields reasonable agreement with $N \approx 5.6$.

In conclusion, our results show that the quantitative accuracy of the 1D model rests on the assumptions made to describe interlayer couplings and internal screening effects.
Many fitting parameters might be required to properly describe more complex systems. 
On the other hand, the model is able to shed light on qualitative trends in the optical activity of multilayer systems, providing an advanced baseline of interpretation without the cost of a full \textit{ab initio} calculation. 

\section{Conclusions}

We have provided a detailed explanation of the splitting of excitonic states in few-layer hexagonal boron nitride. 
Surface effects lead to an energetic separation of excitons localized on the inner layers and excitons localized on the outer layers. 

We have systematically studied the effect of layer number on the electronic and optical propertes of few-layer hexagonal boron nitride sheets. We have presented full GW and BSE calculations of monolayer, bilayer, trilayer, pentalayer and bulk hBN.
Concerning the electronic band structure, we observe that due to enhanced screening with increasing number of layers, the direct gap of hBN decreases from 7.3 eV in single-layer hBN to 6.5 eV in pentalayer and 6.25 eV in bulk hBN. 
At the same time, the excitonic binding energy is reduced such that the center of gravity of the lowest bound exciton remains almost constant. Furthermore, we observe a Davydov splitting of the excitons in a way similar to the splitting of phonon modes with increasing number of layers.
We have analyzed the split excitonic states in terms of energy, localization, symmetry, and optical activity.
In order to elucidate the physical mechanisms of the splitting, we have developed a tight-binding model that is able to efficiently and quite accurately describe excitonic effects in multilayer systems. The ``linear chain'' model for the interlayer interaction provides an analytic formula for the energy splitting of excitonic states within a Davydov multiplet. 
We hope that this work stimulates the ongoing experimental investigations of boron nitride systems, while also being helpful for theoretical studies of the effects of stacking on the optical properties of other layered materials.

\begin{acknowledgments}
F. P., A M.-S. and L. W. acknowledge support from the National Research Fund, Luxembourg (Projects EXCPHON/11280304, C14/MS/773152/FAST-2DMAT and INTER/ANR/13/20/NANOTMD, respectively). We acknowledge fruitful discussions about the visualization of excitonic wave functions with H. Miranda. S. Latil is gratefully acknowledged for providing his tight-binding code.
The research leading to these results has received funding from the European Union H2020 Programme under grant agreement no. 696656 GrapheneCore1.
\end{acknowledgments}

\appendix
\section{Additional computational details}\label{app:cdetails}

The convergence of the internal \texttt{Yambo} parameters was carefully checked by regularly increasing each one until differences in band energies (for GW) or excitonic peak positions (for BSE) were less than $0.01$ eV each time (except for the pentalayer, where the threshold was increased to $0.02$ eV), which is the precision of the GW method. As we are dealing with quasi-2D materials, special attention was paid to the amount of vacuum space introduced between repeated copies of the systems in the vertical direction. Because of the long-range Coulomb tail of the response functions that describe the screening, the repeated copies will interact with each other even with a very high separation distance ($d= 40$ \AA).\cite{Wirtz2006} By using a cutoff of the Coulomb interaction in the vertical direction,\cite{Rozzi2006} we were able to obtain converged results with a separation distance $d = 20$ \AA. Another important observation is that in this situation, the convergence of the results with respect to \textit{both} (i) the k-point sampling and (ii) the number of included unoccupied states depends on the size of the supercell.\cite{PhysRevB.93.235435} If $h$ is the thickness of the system and $L_z=h+d$ the supercell height, as we increase the number of atomic layers $L_z$ becomes larger, and consequently we might need to use a denser k-point mesh and to sum over more unoccupied states. 
Table \ref{t:conv} summarizes the parameters to obtain converged GW $\pi$ and $\pi^*$ bands and converged (lowest-lying) excitonic peaks.

In the GW case, the plasmon-pole approximation was used for the computation of the electronic response function.\cite{Larson2013} Its validity was checked, for the monolayer, against the direct integration in frequency space, yielding excellent agreement. Moreover, our GW bandgap value (7.26 eV) for the hBN monolayer is in good agreement with other results obtained with different many-body codes ($7.36$ eV\cite{Cudazzo2016} and $7.37$ eV\cite{Huser2013}). 
The numerical shift of $0.1$ eV is entirely due to the underlying DFT calculation: the cited results can be obtained exactly by switching to the optimized lattice constant for the monolayer.
Our optical spectrum for the monolayer also agrees with the one in Ref. [\onlinecite{Cudazzo2016}]. 

An additional convergence check was performed on the monolayer, by decreasing the convergence threshold by almost an order of magnitude (using a $48\times48\times1$ k-point mesh, a vacuum separation of $30$ \AA, and summing up to $400$ states). 
The results for GW band gap and excitonic peak positions differ by about $0.03$ eV (rigid shift) from the ones obtained with the parameters listed in Tab. \ref{t:conv}. We conclude therefore that our results are well converged. Our reference calculations for the bulk system are in agreement with previously established results.\cite{Arnaud2006,Wirtz2008}

\section{Transition energy region for multi-layer hBN}\label{app:transition}

\begin{figure}
\includegraphics[width=0.5\textwidth]{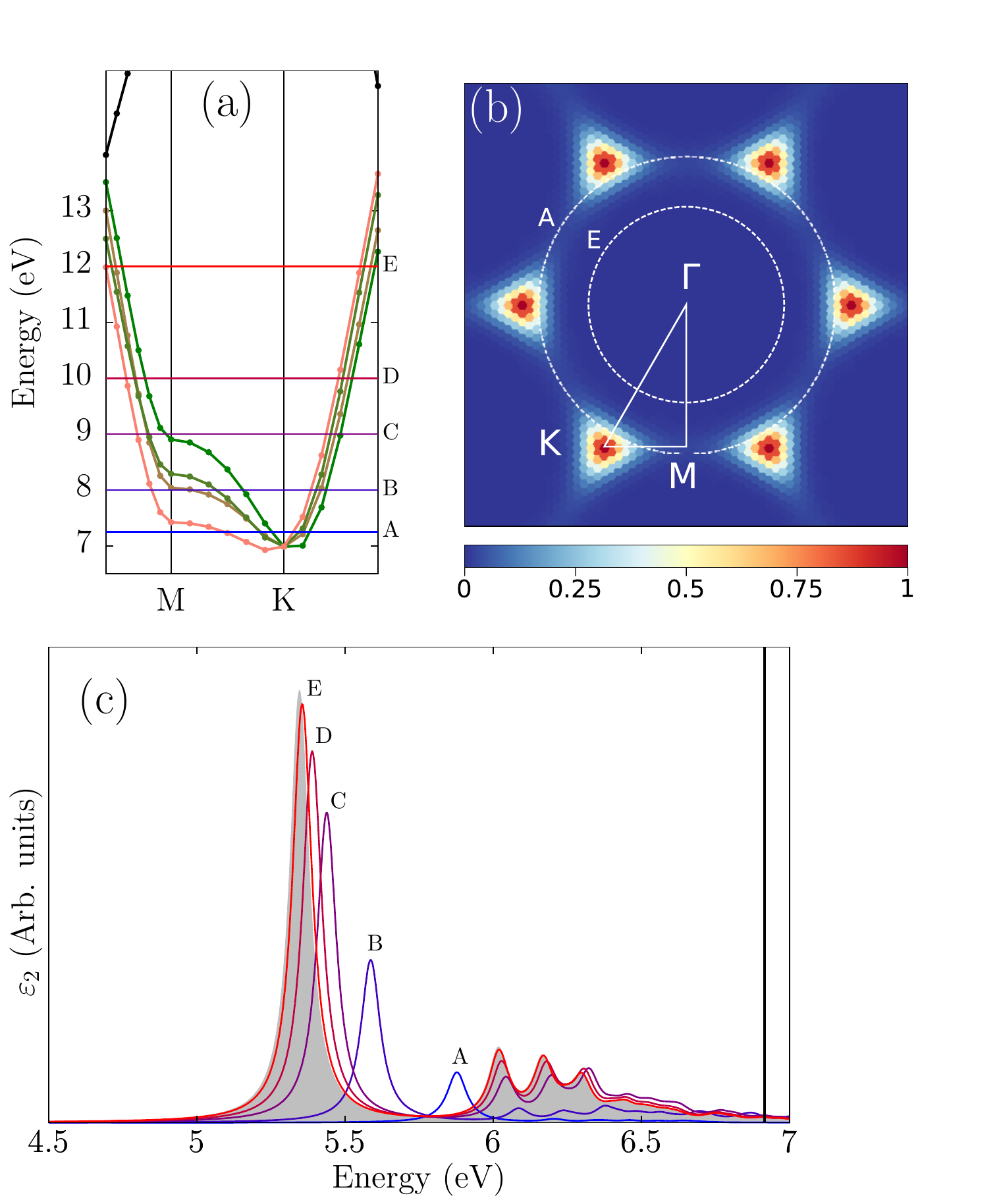}
\caption{Transition energy region (TER) for absorption, in the case of bilayer hBN. In (a), the transition energies obtained from the disentangled $\pi$ and $\pi^*$ bands are shown with different colors in the relevant part of the BZ. The lowest $\sigma \rightarrow \sigma^*$ transition is shown in black. The horizontal lines labeled from A to E represent different TERs. They include (A) only the K point, (B) the lowest transition along KM, (C) all transitions along KM, (D) parts of the $\Gamma$M and $\Gamma$K directions, (E) all energies of the relevant region. The excitonic weights in $k$-space for the first bright exciton are shown in (b). Most of the weight comes from the area around the K point. The intersections between the irreducible wedge of the BZ (white triangle) and the white dashed circles labeled A,E represent the fraction of the BZ which is included in the BSE calculations in the two cases. The imaginary part of the dielectric functions obtained using the five TERs from A to E are shown in (c). It can be seen how only a very wide TER -- in this case the one labeled (E) -- is able to reproduce the fully converged result (gray shadow).}\label{f:transitions}
\end{figure}

Let us consider bilayer hBN. 
In Fig. \ref{f:transitions}(a), The transition energies $\Delta_{cv}(\mathbf{k})=E_c(\mathbf{k})-E_v(\mathbf{k})$ obtained from the disentangled GW valence and conduction bands are shown in different colors. 
In order to obtain converged \textit{ab initio} results for the absorption spectra in multilayer hBN, one might be tempted to only include in the calculations the area around the K point or along the KM region in the BZ (transitions below lines (A), (B) or (C) in the figure).
This seems justified by looking at Fig. \ref{f:transitions}(b), which shows the weights -- i.e. the Fourier intensities $\sum_{cv}\Psi^\lambda_{cv}(\mathbf{k})$ -- of the electronic transitions in the BZ for the lowest-bound bright exciton. 
However, it can be seen from Fig. \ref{f:transitions}(c) that this would produce unconverged spectra. 
The converged result is obtained by increasing the energy window included in the calculation up to the $\pi^*$-$\sigma^*$ crossing ((E) lines in Fig. \ref{f:transitions}(a)-(b)). 

\section{Derivation of the tight-binding model}
\label{app:TB} 
The wide band gap of hBN justifies the approximation in which the TB Hamiltonian is separated into an effective low-energy part $\hat{H}_h$ acting on the holes and an effective high-energy part $\hat{H}_e$ acting on the electrons. We describe the space of states of the system as a tensor product of hole and electron states, and thus rewrite our kinetic/single-particle Hamiltonian as $\hat{H}_{0} \approx \mathbb{Id}_{h} \otimes \hat{H}_{e} - \hat{H}_h \otimes \mathbb{Id}_e$.
The effective Hamiltonians are obtained from second order perturbation theory by considering the hopping terms $\tpar$ and $\tper$ as perturbations of the system:
\begin{multline*}
\hat{H}_{h}  \approx  - \sum_{\vec{m} \in \Lambda_h} \qty(\Delta +\frac{\tpar^2}{2 \Delta} \mathcal{N}_\parallel \qty(\vec{m})+\frac{\tper^2}{2 \Delta} \mathcal{N}_\perp \qty(\vec{m})) \ket{\alpha, \vec{m}}\bra{\alpha, \vec{m}} \\ - \sum_{<\vec{m}, \vec{m'}>_{\parallel}} \frac{\tpar^2}{2 \Delta} \ket{\alpha, \vec{m}}\bra{\alpha, \vec{m'}} - \sum_{<\vec{m}, \vec{m'}>_{\perp}} \frac{\tpar \tper}{2 \Delta} \ket{\alpha, \vec{m}}\bra{\alpha', \vec{m'}}
\end{multline*}
\begin{multline*}
\hat{H}_{e}  \approx  \sum_{\vec{n} \in \Lambda_e} \qty(\Delta +\frac{\tpar^2}{2 \Delta} \mathcal{N}_\parallel \qty(\vec{n})+\frac{\tper^2}{2 \Delta} \mathcal{N}_\perp \qty(\vec{n})) \ket{\beta, \vec{n}}\bra{\beta, \vec{n}} \\ + \sum_{<\vec{n}, \vec{n'}>_{\parallel}} \frac{\tpar^2}{2 \Delta} \ket{\beta, \vec{n}}\bra{\beta, \vec{n'}} + \sum_{<\vec{n}, \vec{n'}>_{\perp}} \frac{\tpar \tper}{2 \Delta} \ket{\beta, \vec{n}}\bra{\beta', \vec{n'}} ,
\end{multline*}
where $<\cdot, \cdot>_\parallel$ ($<\cdot, \cdot>_\perp$) denotes summation over in-plane (out-of-plane) nearest neighbors of the same species and $\mathcal{N}_\parallel \qty(\cdot)$ ($\mathcal{N}_\perp \qty(\cdot)$) denotes the number of in-plane (out-of-plane) nearest neighbors of the opposite species (boron for nitrogen and nitrogen for boron). The integers $\mathcal{N}_\parallel \qty(\cdot)$ and $\mathcal{N}_\perp \qty(\cdot)$ depend on the geometry of the system, and thus on the stacking of the layers. In the case of the AA$^\prime$ stacking, we get $\mathcal{N}_\parallel \qty(\vec{n}) = 3$ and:
\begin{equation*}
\mathcal{N}_\perp \qty(\vec{n}) = 
\begin{cases} 
2 &\mbox{if } \vec{n} \mbox{ is in the inner layers} \\ 
1 &\mbox{if } \vec{n} \mbox{ is in the outer layers}
\end{cases}.
\end{equation*}

Let us now consider the excitonic basis and Hamiltonian defined in Eqs. \eqref{eq:TBbasis} and \eqref{eq:TBham}, respectively.
As mentioned in Section \ref{s:TB}, one of the main advantages of the basis of direct-space excitations $\{\ket{\vec{R}_{\alpha, \beta}}\}$ with well-defined electron-hole vectors is that Coulomb matrix-type elements are easily expressed in it. Indeed,to lowest order, the direct interaction is diagonal:\cite{monolayer}
\[
\bra{\vec{R}_{\alpha, \beta}}\hat{U}\ket{\vec{R'}_{\alpha ', \beta '}} \approx \delta_{\vec{R}_{\alpha, \beta}, \vec{R'}_{\alpha ', \beta '}} U_{\vec{R}_{\alpha, \beta}},
\]
where the quantities $U_{\vec{R}_{\alpha, \beta}}$ can be approximated by a model electron hole potential $V_{\qty(\alpha, \beta)}\qty(\vec{R})$. 
The simplest  model potential would be a simple screened Coulomb potential, but it has already been pointed out\cite{monolayer, Wu2015, Cudazzo2011, Chernikov2014, Pulci2015, Berkelbach2013, Rodin2014, Latini2015} that it is not suitable for the description of anisotropically screened 2D systems.  Here, we will make use of a modified Keldysh potential.\cite{Cudazzo2011,Keldysh1979} Having obtained the matrix elements of the electron-hole interaction, what remains to be done is to calculate  the matrix elements of $\hat{H}_{0}$ in the $\ket{R_{\alpha, \beta}}$ basis.  This is readily done by inserting the definition of $\ket{R_{\alpha, \beta}}$, and recalling that $\hat{H}_{0} = \idop_{h} \otimes \hat{H}_{e} - \hat{H}_h \otimes \idop_e$. We find:
\begin{equation*}
\begin{split}
\bra{\vec{R}_{\alpha, \beta}}\hat{H}_0 \ket{\vec{R'}_{\alpha ', \beta '}} = \frac{\delta_{\alpha, \alpha '}}{M} \sum_{\vec{n} \in {\Lambda}_{e, \beta}} \bra{\beta, \vec{n}}\hat{H}_e\ket{\beta ', \vec{n}+\vec{R'}-\vec{R}} \\ - \frac{\delta_{\beta, \beta '}}{M} \sum_{\vec{m} \in {\Lambda}_{h, \alpha}}\bra{\alpha, \vec{m}}\hat{H}_h\ket{\alpha ', \vec{m+\vec{R}-\vec{R'}}} .
\end{split}
\end{equation*}
Therefore, the matrix elements of the kinetic Hamiltonian are derived from those of the effective Hamiltonian. Since these are stacking dependent, we will now specialize to the AA$^\prime$ case and use the previously derived expressions. We thus get, for the diagonal elements:
\begin{equation*}
\bra{\vec{R}_{\alpha, \beta}}\hat{H}_0 \ket{\vec{R}_{\alpha, \beta}} = 2\Delta + 3\frac{\tpar^2}{\Delta} + \frac{\mathcal{B}\qty(\alpha, \beta)}{2}\frac{\tper^2}{\Delta} ,\\
\end{equation*}
and for the non diagonal elements:
\begin{multline*}
\bra{\vec{R}_{\alpha, \beta}}\hat{H}_0 \ket{\vec{R'}_{\alpha ', \beta '}} = \\
\begin{cases}
\frac{\tpar^2}{\Delta} \quad &\mbox{if $\vec{R}$ and $\vec{R'}$ are 1.n.n. and $\alpha=\alpha '$ and $\beta = \beta '$} \\
\frac{\tpar\tper}{\Delta} &\mbox{if $\vec{R}$ and $\vec{R'}$ are 1.n.n. and $\abs{\alpha-\alpha '}+\abs{\beta - \beta '}=1$} \\
0 &\mbox{otherwise},
\end{cases}
\end{multline*}
where the quantity $\mathcal{B}\qty(\alpha, \beta)$ is given, in the AA$^\prime$ stacking by:
\begin{equation*}
\mathcal{B}\qty(\alpha, \beta) =
\begin{cases}
2 &\mbox{if $\alpha,\beta \in \qty{1,N}$} \\
3 &\mbox{if $\alpha \in \qty{1,N}$ and $\beta \in \llbracket 2, N-1 \rrbracket$ } \\
3 &\mbox{if $\beta \in \qty{1,N}$ and $\alpha \in \llbracket 2, N-1 \rrbracket$ } \\
4 &\mbox{if $\alpha,\beta \in \llbracket 2, N-1 \rrbracket$}
\end{cases}
\end{equation*}
The quantity $\mathcal{B}\qty(\alpha, \beta)$  has a physical meaning: if we extend the notation $\mathcal{N}_\perp \qty(\cdot)$ by noticing that all hole (electron) sites in a given layer $\alpha$ ($\beta$) have the same number of electron (holes) out-of-plane nearest neighbors $\mathcal{N}_\perp^{(h)} \qty(\alpha)$ ($\mathcal{N}_\perp^{(e)} \qty(\beta)$), we have that:
\[
\mathcal{B}\qty(\alpha, \beta) = \mathcal{N}_\perp^{(h)} \qty(\alpha) + \mathcal{N}_\perp^{(e)} \qty(\beta).
\]
In other words, $\mathcal{B}\qty(\alpha, \beta)$ counts the out-of-plane ``coordination number" of the sites taking part in the excitation, each out-of-plane nearest neighbor contributing an energy of $\frac{\tper^2}{2\Delta}$ to the kinetic energy of the excitation. In the case of in-plane bonds the number of in-plane nearest neighbors is the same for all sites, since the layers all have the same structure: each site has $\mathcal{N}_\parallel = 3$ nearest neighbors, leading to a $\mathcal{B}_\parallel = 3 + 3 = 6$. Each in-plane nearest neighbor contributes an energy of $\frac{\tpar^2}{2\Delta}$ to the kinetic energy of the excitation, yielding a total contribution of $3\frac{\tpar^2}{\Delta}$, as can be seen in the formulas above. The fact that $\mathcal{B}\qty(\alpha, \beta)$ is not constant is a consequence of the finite number of layers and therefore sites in the outer layers have less nearest neighbors than sites in the inner layers. As a result,  excitations involving the outer layers have less kinetic energy than excitations involving the inner layers. This will have consequences on the splitting of the excitonic states.

\section{Lattice of direct space excitations}
\label{app:exclat}
The Hamiltonian of Eq. \eqref{eq:TBham2} can be interpreted in geometrical terms. Each element of the basis $\{\ket{\vec{R}_{\alpha, \beta}}\}$ is associated to a point at position $\vec{R}$ denoted by $\qty(\alpha, \beta)$ and called excitation site, $\vec{R}_{\alpha, \beta}$. The set of  excitation sites is a set of discrete points in the geometric space, in the same way as electronic sites in electronic tight-binding models. 

The set of excitation sites inherits a lattice structure from the physical lattice of the hBN multilayer. Recall that the vectors $\vec{R}$ range over the possible electron-hole vectors allowed in the physical lattice. In the single layer case, the set of such vectors is a triangular lattice with the origin chosen at the center of one triangle, and then attaching to each excitation site the corresponding amplitude $\braket{\vec{R}_{\alpha, \beta}}{\Psi}$ of the excitonic state yields the usual fixed-hole representation of excitonic states in direct space. This is because, in the monolayer, all lattice positions of the hole are equivalent. In multilayers, this is no longer the case: while it is still true that all positions of the hole within a given layer are equivalent, the layers are inequivalent, so one has to sweep the position of the hole (nitrogen atoms) over all layers in order to reconstruct the full symmetry of the wave function.

The lattice of excitations is constructed with a general procedure: for each couple of layers $\qty(\alpha, \beta)$, select one hole position in layer $\alpha$ (the exact position chosen does not matter, as all hole positions \textit{within that layer} are equivalent), and then consider all the electron hole vectors from this position of the hole to the possible electron positions (boron atoms) of layer $\beta$. 
We obtain in this way a set of vectors $\mathcal{L}_{\alpha, \beta}$.  Note that the sets $\mathcal{L}_{\alpha, \beta}$ for different $\qty(\alpha, \beta)$ are not necessarily disjoint: a given hole-electron vector can be realized in several pairs of layers, and so different excitation sites might have the same position in the lattice. This, along with notational convenience, is the reason why excitation sites must be labeled by an $\qty(\alpha, \beta)$ index. We thus naturally define the \textit{excitation sublattices} as
\begin{equation*}
\Lambda_{\alpha, \beta}=\qty{\vec{R}_{\alpha, \beta} \ | \ \vec{R} \in \mathcal{L}_{\alpha, \beta}},
\end{equation*}
where $\Lambda_{\alpha, \beta}$ is obtained by taking all points of $\mathcal{L}_{\alpha, \beta}$ and labeling them with the indices $(\alpha, \beta)$. The whole lattice of excitations is then nothing but the union of all excitation sublattices $\Lambda_{\alpha, \beta}$. The sublattices have physical meaning: they are the set of direct space exitations with the hole in layer $\alpha$ and the electron in layer $\beta$.

We can use the method described above to obtain the excitation sublattices explicitly in the case of the $AA'$ stacking. Let $\mathcal{T}$ denote the triangular lattice defined by the electron sites / boron centers in layer $1$ and $\vec{\tau}$ be a first nearest neighbour nitrogen-boron vector in this layer. Let also $d$ be the interlayer distance, and $\vec{e}_z$ be a unit vector along the stacking direction. We find:
\begin{multline*}
\Lambda_{\alpha, \beta}=\Bigg\{\vec{R}_{\alpha, \beta} \ \bigg| \ \vec{R} \in \mathcal{T} + {\qty(-1)}^{\alpha-1}\frac{1+\qty(-1)^{\beta-\alpha}}{2} \vec{\tau}  \\ + \qty(\beta-\alpha) d \vec{e}_z \Bigg\}.
\end{multline*}
In particular, $\alpha = \beta$ corresponds to excitations confined in one given layer (in-plane sublattcies). On the other hand, sublattices with $\alpha \neq \beta$ correspond to excitations with hole and the electron in different layers (interlayer sublattices).

Moreover sublattices are geometrically equivalent if
\begin{equation*}
\Lambda_{\alpha, \beta} \sim \Lambda_{\alpha ', \beta '} \iff \qty(\mathcal{L}_{\alpha, \beta} = \mathcal{L}_{\alpha ', \beta '} \text{ or } \mathcal{L}_{\alpha, \beta}=\hat{I} \qty(\mathcal{L}_{\alpha ', \beta '})),
\end{equation*}
where $\hat{I}$ denotes the inversion symmetry and $\sim$ marks the equivalence relation and equivalent sublattices fulfill:
\begin{equation*}
\Lambda_{\alpha, \beta} \sim \Lambda_{\alpha ', \beta '} \iff \abs{\alpha - \beta}=\abs{\alpha ' - \beta '}.
\end{equation*}

Therefore, in a $N$-layer system there are only $N$ equivalent classes for the sublattices. The coupling between sublattices is governed by the effective hopping term $T_\perp = \frac{\tpar \tper}{\Delta}$. In addition, the potential terms $V_{\alpha, \beta}$ do not vary strongly within one class and at zeroth-order all sublattices decouple and the Hamiltonian is block diagonal, with each 
block corresponding to geometrically equivalent sublattices. Conceptually, this means that we can obtain a good approximation of the splitting behavior of the $N$-layer system by studying one sublattices per equivalence class and then use perturbation theory to study the behavior of the full system, as governed by the kinetic coupling.

The sublattices are useful to analyse the eigenstates of the Hamiltonian and to provide approximate methods of diagonalization. The kinetic Hamiltonian $\hat{H}_0$ describes two types of hoppings: hoppings between nearest neighbour excitations within the same sublattice, with hopping amplitude $T_\perp = \frac{{t_\perp}^2}{\Delta}$, and hoppings between different sublattices with amplitude $T_\parallel = \frac{t_\perp t_\parallel}{\Delta}$. These are only possible if the index corresponding to the hole ($\alpha$) or the index corresponding to the electron ($\beta$), but not both, change by exactly $1$. This corresponds to the physical situation where either the hole or the electron effectively jumps from one nitrogen / boron (resp.) site in a layer to a nitrogen / boron (resp.) site in a neighbouring layer.

\section{Derivation of the multilayer effective Hamiltonian}\label{app:1D}

We derive the effective Hamiltonian $\hat{H}_{\text{eff}}$ for the description of the lowest-bound Davydov multiplet in $N$-layer systems. All sublattices $\Lambda_{\alpha, \beta}$ with the same $\abs{\beta-\alpha}$ are geometrically equivalent as stated above. In the absence of relevant screening variations, geometrically equivalent sublattices have the same interaction potential $V_{\qty(\alpha, \beta)}$, which depends only on $\abs{\beta - \alpha}$.

We define a Hamiltonian $\bar{H}_\parallel$ where the functions $V_{\qty(\alpha, \beta)}$ have been replaced by their averages $\bar{V}_{\beta-\alpha}$.  Correspondingly, we have isolated the variations in screening $\hat{U}_Z= \Hpar - \bar{H}_\parallel$. By construction, $\bar{H}_\parallel$ describes the problem of a set of non-interacting sublattices with the same $\beta-\alpha$.
Physically, it is a Hamiltonian for a collection of effective \textit{identical} monolayers whose electrons and holes cannot hop between layers. In particular, two adjacent layers are thus symmetric under their inversion. This increased symmetry allows us to build an eigenbasis $\mathcal{B}_0$ of $\bar{H}_\parallel$ in the same way as it was done for the bilayer. In particular, the ground state of $\bar{H}_\parallel$ is associated to a $2N$ dimensional eigensubspace spanned by $N$ copies of a monolayer ground state with modified screening. In the end, we have thus decomposed the excitonic Hamiltonian:
\begin{equation*}
\hat{H}_X = \bar{H}_\parallel + \Hper + \hat{U}_Z ,
\end{equation*}
so that the splitting effects are described by the operator $\hat{H}_1 = \Hper + \hat{U}_Z$, which we will treat as a perturbation of the problem described by $\bar{H}_\parallel$. To this end, we require the matrix elements of $\Hper$ and $\hat{U}_Z$ in the basis $\mathcal{B}_0 = \qty{\ket{\Psi_i}}_{i \in \mathbb{N}}$. 
Since the matrix elements of these operators are known in the basis of excitations $\vec{R}_{\alpha, \beta}$, their matrix elements in $\mathcal{B}_0$ are obtained by expanding the elements of $\mathcal{B}_0$ in the basis $\vec{R}_{\alpha, \beta}$:
\begin{equation*}
\ket{\Psi_i} = \sum_{\vec{R}_{\alpha, \beta}}  \Psi_{i, \vec{R}_{\alpha, \beta}} \ket{\vec{R}_{\alpha, \beta}} .
\end{equation*}
For $\Hper$, in the case of the $AA^\prime$ stacking, we obtain:
\begin{equation*}
\bra{\Psi_i}\Hper\ket{\Psi_j} = \delta_{i, j} \frac{\mathcal{B}\qty(\alpha_i, \beta_i)}{2} \frac{{t_\parallel}^2}{\Delta} + s_{i,j} \frac{t_{\perp}t_{\parallel}}{\Delta} ,
\end{equation*}
where:
\begin{equation*}
s_{i, j} =  \sum_{\vec{R}_{\alpha_i, \beta_i}} \sideset{}{'}\sum_{\vec{R}^\prime_{\alpha_j, \beta_j}} \Psi_{i, \vec{R}_{\alpha_i, \beta_i}}^* \Psi_{j, \vec{R}^\prime_{\alpha_j, \beta_j}} ,
\end{equation*}
with the primed sum extending over the set of the out-of-plane nearest neighbors of $\vec{R}_{\alpha_i, \beta_i}$ with non-zero hopping elements; or, in other words, the sets of its nearest neighbors $\vec{R}_{\alpha_j, \beta_j}$ such that $\abs{\alpha_j-\alpha_i}+\abs{\beta_j - \beta_i}=1$. Note that, as a result, if $\abs{\alpha_j-\alpha_i}+\abs{\beta_j - \beta_i} \neq 1$, then $s_{i,j}=0$. For later convenience, we will say that sublattices $\Lambda_{\alpha_i, \beta_i}$ and $\Lambda_{\alpha_j, \beta_j}$ are \textit{connected} when the condition $\abs{\alpha_j-\alpha_i}+\abs{\beta_j - \beta_i}=1$ is met. For $\hat{U}_Z$, in the case of the $AA^\prime$ stacking, since $\hat{U}_Z$ is diagonal (because $\hat{U}$ is diagonal), we obtain:
\begin{equation*}
\bra{\Psi_i}\hat{U}_Z\ket{\Psi_j} = u_{i,j} ,
\end{equation*}
where:
\begin{equation*}
u_{i,j} = \sum_{\vec{R}_{\alpha_i, \beta_i}} \Psi_{i, \vec{R}_{\alpha_i, \beta_i}}^* \Psi_{j, \vec{R}_{\alpha_i, \beta_i}} \bra{\vec{R}_{\alpha_i, \beta_i}}\hat{U}_Z\ket{\vec{R}_{\alpha_i, \beta_i}}
\end{equation*}
which shows that, as expected, $\hat{U}_Z$ does not couple states from different sublattices, so that one may also write: $\bra{\Psi_i}\hat{U}_Z\ket{\Psi_j} = \delta_{(\alpha_j, \beta_i),  (\alpha_j, \beta_j)} u_{i,j}$. 
Notice in particular that, as a result, $u_{i,j}$ and $s_{i,j}$ cannot both be non-zero at the same time.

We can now write the form of the effective Hamiltonian for the splitting. We will consider here only the splitting of the lowest-lying exciton, since it is expected to contribute the most to the absorption spectra of hBN, and because its associated eigensubspace is well separated in energy from the other states, which is a necessary condition for accurate degenerate perturbation theory.  Let us thus use the same procedure as in the case of the bilayer: for each layer, we consider one effective copy of the monolayer ground state exciton so that these states are all images of each other by inversion symmetry of $\bar{H}$. We denote this set of $N$ uncoupled states as ${\qty{\ket{i}}}_{i \in {\llbracket 1,N \rrbracket}}$ where $i$ now labels  the layer  and varies from $1$ to $N$. The corresponding effective Hamiltonian, up to second order is thus given by (the zeroth order part is shifted away):
\begin{equation*}
\bra{i}\hat{H}_{eff}\ket{j} =\bra{i}\hat{H}_1\ket{j}  + \sum_{\mu} \frac{\bra{i}\hat{H}_1\ket{\mu}\bra{\mu}\hat{H}_1\ket{j}}{E_D - E_{\mu}}
\end{equation*}
where $E_D$ is the eigenenergy associated with the degenerate subspace formed by the first monolayer excitons and $\qty{\ket{\mu}}$ is the set of elements of $\mathcal{B}_0$ outside of that subspace.

The first order terms are readily obtained: since all states in ${\qty{\ket{i}}}_{i \in {\llbracket 1,N \rrbracket}}$ are on different sublattices that are not connected to each other, $\hat{H}_1$ is diagonal in this basis, hence:
\[
{\hat{H}_{eff}}^{(1)} = \sum_{i=1}^N \qty(\frac{\mathcal{B}\qty(\alpha_i, \beta_i)}{2}\frac{\tper^2}{\Delta} + u_{i,i}) \dyad{i}{i}
\]
Since $\hat{H}_1 = \Hper + \hat{U}_Z$, the second order terms result \textit{a priori} in three types of terms: quadratic terms in $\Hper$, quadratic terms in $\hat{U}_Z$ and cross terms. 
Since $u_{i,j}$ and $s_{i,j}$ are never both non-zero, however, the cross terms vanish, and we are left only with the quadratic terms. 
Again, since all elements of the set ${\qty{\ket{i}}}_{i \in {\llbracket 1,N \rrbracket}}$ are from different sublattices, the quadratic terms in $\hat{U}_Z$ must be diagonal. 
From the form of their matrix elements, the quadratic terms in $\Hper$ can only be nonzero for a certain pair $\qty(\ket{i},\ket{j})$ if there exists some state $\ket{\mu}$ such that the sublattice of $\ket{\mu}$ is connected to the sublattices of $\ket{i}$ and $\ket{j}$, so that a coupling is only possible if $\abs{i-j} \leq 1$. 
As a result, these terms are tridiagonal in ${\qty{\ket{i}}}_{i \in {\llbracket 1,N \rrbracket}}$.

For convenience, we introduce the notation $\mathcal{C}_{i,j}$ to denote the set of sublattices that are connected to both $\Lambda_{i,i}$ and $\Lambda_{j,j}$. 
Making now use of this allows us to express $\hat{H}_{\text{eff}}$ in the form:
\begin{widetext}
\begin{multline*}
\hat{H}_{\text{eff}} = \sum_{i=1}^N \qty[\frac{\mathcal{B}\qty(\alpha_i, \beta_i)}{2}\frac{\tper^2}{\Delta} + u_{i,i} + \sum_{\mu \in \Lambda_{i,i}} \frac{\abs{u_{i,\mu}}^2}{E_D - E_{\mu}} + {\qty(\frac{t_\perp t_\parallel}{\Delta})}^2 \sum_{\mu \in \mathcal{C}_{i,i}} \frac{\abs{s_{i, \mu}}^2}{E_D - E_{\mu}}] \dyad{i}{i} \\ + \sum_{<i,j>}\qty[ {\qty(\frac{\tpar \tper}{\Delta})}^2 \sum_{\mu \in \mathcal{C}_{i,j}} \frac{{s_{i, \mu}}^* s_{\mu , j}}{E_D - E_{\mu}}]\dyad{i}{j}
\end{multline*}
\end{widetext}

Let us now make use of the symmetries of the states in $\mathcal{B}_0$: since its states are chosen according to the symmetry under inversion of two adjacent layers, it follows that for any sublattice $\Lambda \in \mathcal{C}_{i,j}$ with $\abs{i-j} \leq 1$, the quantity ${\qty(\frac{\tpar \tper}{\Delta})}^2 \sum_{\mu \in \Lambda} \frac{{s_{i, \mu}}^* s_{\mu , j}}{E_D - E_{\mu}}$ can only take two values.
Indeed, if $i=j$, then  ${s_{i, \mu}} = s_{\mu , j}$ and the sum has some value $\frac{h}{2} \leq 0$. If $i \neq j$, ${s_{i, \mu}} \neq s_{\mu , j}$ in general because the layers $i$ and $j$ have reversed orientation. In this case, we call $\frac{g}{2}$ the value of the sum.
It follows that the value of a sum of the form ${\qty(\frac{\tpar \tper}{\Delta})}^2 \sum_{\mu \in \mathcal{C}_{i,j}} \frac{{s_{i, \mu}}^* s_{\mu , j}}{E_D - E_{\mu}}$ is simply $\frac{h}{2}$ or $\frac{g}{2}$ times the number of sublattices in the set $\mathcal{C}_{i,j}$. 
If $i=j$ and $i \in {\llbracket 2,N-1 \rrbracket}$ there are four of them ($\Lambda_{i, i+1}$, $\Lambda_{i+1, i}$, $\Lambda_{i-1, i}$ and $\Lambda_{i, i-1}$), two of them if $i=j$ and $i =1 \mbox{ or } N$ ($i+1$ or $i-1$ is not in ${\llbracket 1,N \rrbracket}$ then) and two if $\abs{i-j}=1$ ($\Lambda_{i, j}$ and $\Lambda_{j, i}$).

Another simplifying remark can be made: the values of the $\mathcal{B}\qty(\alpha_i, \beta_i)$ are known (see appendix \ref{app:TB}): $4$ if $i \in {\llbracket 2,N-1 \rrbracket}$ and $2$ if $i =1 \mbox{ or } N$. 
Thus, if we perform a
shift the energy scale by $-2\frac{\tper^2}{\Delta}-2h$, and give a name to the quantities related to the variations in screening: $u_i = u_{i,i}+\sum_{\mu \in \Lambda_{i,i}} \frac{\abs{u_{i,\mu}}^2}{E_D - E_{\mu}}$, this leaves us with:
\begin{multline*}
\hat{H}_{eff} = g \sum_{<i,j>}\dyad{i}{j} + \qty(-\frac{\tper^2}{\Delta}+\abs{h}) \qty(\dyad{1}+\dyad{N})  \\ + \sum_{i=1}^N u_i \dyad{i}{i} .
\end{multline*}

\textit{A priori}, the sign of $g$ is not known. However, from the above Hamiltonian, we can see that $g$ corresponds to an interlayer coupling term: it is the multilayer analogue of the bilayer quantity $g_\Psi$. From section \ref{ss:mod2L}, it is known that $g_\Psi \leq 0$ from the lowest bound Davydov pair of the bilayer, and \textit{ab initio} calculations of 3 and 5 layers systems (see section \ref{s:gaps}) indicate that $g$ remains negative in these cases, and seems to be independent of $N$. We therefore take $g < 0$ for all $N$, and write $g\equiv -\abs{g}$ from now on.

At this point, in order to obtain a simple model, we can make the approximation that the variations of the screening along the stacking direction can be effectively modeled by considering this variation only on the outer layers $1$ and $N$. 
In other words, we suppose that, up to a shift of the energy scale, there is a real $u$ such that:
\begin{equation*}
u_i \approx \qty(\delta_{i, 1} + \delta_{i, N}) \ u ,
\end{equation*}
so that the effective Hamiltonian reduces to the problem of a linear chain with border effects:
\begin{equation*}
\hat{H}_{eff} = -  \abs{g} \sum_{<i,j>}\dyad{i}{j} + \qty(-\frac{\tper^2}{\Delta}+u+\abs{h}) \qty(\dyad{1}+\dyad{N}) .
\end{equation*}
Defining now the dimensionless parameter $X=\frac{1}{\abs{g}}\qty(\frac{\tper^2}{\Delta }-u-\abs{h})$ as the ratio between the border terms and the hopping terms, we can rewrite the effective Hamiltonian into the following form:
\begin{equation*}
\hat{H}_{eff} = -  \abs{g}\qty[ \sum_{<i,j>}\dyad{i}{j} + X \big(\dyad{1}+\dyad{N}\big)] .
\end{equation*}

\section{Bulk limit in the linear chain model}\label{app:bulk}

For completeness, we provide here the bulk case in the linear chain formalism. In this case, the chain is infinite, and we label the layers with relative integers. The corresponding Hamiltonian is given by:
\begin{equation*}
\hat{H}_{eff} = -  \abs{g}\sum_{<n,m>}\dyad{n}{m}
\end{equation*}
where there is no border term in $X$ because there are no borders. This infinite linear chain is well known: its eigenvalues and eigenvectors can be labeled by some $k\in\qty[-\pi, \pi]$ and are given by:
\begin{equation*}
\ket{k} = \frac{1}{\sqrt{N}}\sum_{n \in \mathbb{Z}} e^{ikn} \ket{n} \quad ; \quad
E\qty(k) = -2\abs{g} \cos\qty(k).
\end{equation*}
The real periodicity of bulk $AA'$ along the stacking direction is two layers, so this exciton band structure must be folded.

We are interested here only in direct excitons, so only in states at the excitonic $\Gamma$ point of this folded band structure, which is to say $\ket{0}$ and $\ket{\pi}$:
\begin{alignat*}{2}
&\ket{0} = \frac{1}{\sqrt{N}}\sum_{n \in \mathbb{Z}} \ket{n}  &&; \quad
E\qty(0) = -2\abs{g} \\
&\ket{\pi} = \frac{1}{\sqrt{N}}\sum_{n \in \mathbb{Z}} {\qty(-1)}^n \ket{n}  \quad &&; \quad
E\qty(\pi) = 2\abs{g}
\end{alignat*}
So, as is known,\cite{Koskelo2017} we recover a splitting in an even ($\ket{0}$) and an odd ($\ket{\pi}$) state, with the even one being the lowest in energy and a Davydov splitting of $s_{bulk}=4\abs{g}$. In bilayer $AA'$, it was found that $\abs{g} \approx 15$ meV, so we expect the splitting in bulk to be about twice that of the bilayer, at $s_{bulk}\approx 60$ meV, in very good agreement with the ab-initio value of $58$ meV. It is easily shown that, with proper normalization, $S\qty(\ket{k}) = \delta_{k, \pi}$ so that $\ket{0}$ is dark and $\ket{\pi}$ is bright, as expected. Other states ($k \neq 0 \text{ or } \pi$) are indirect, and therefore dark.

\section{Phase plot of degenerate excitons from \textit{ab initio}}\label{app:phase}

\begin{figure}
\includegraphics[width=0.5\textwidth, trim={0cm 1.3cm 0cm 0cm},clip]{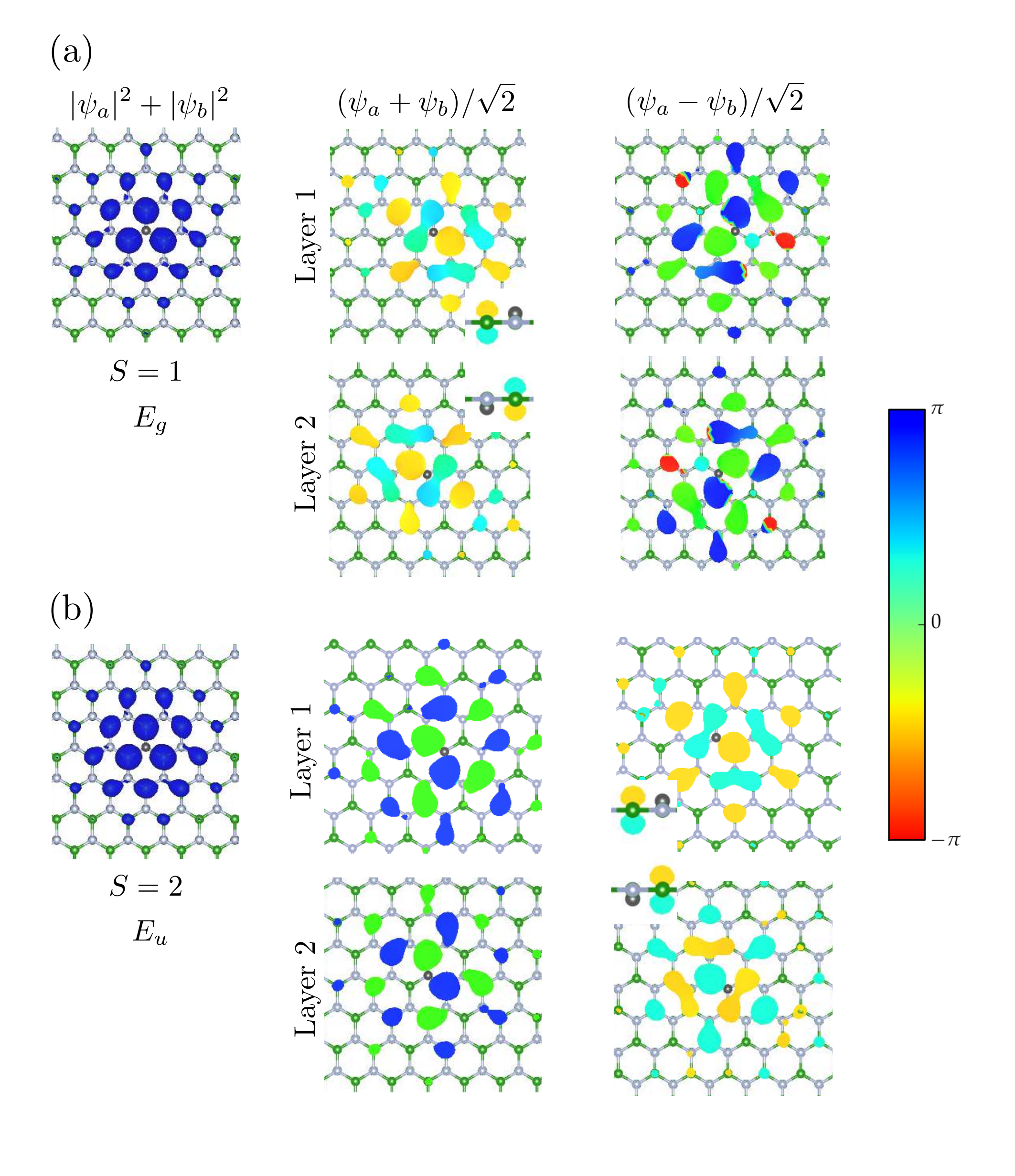}
\caption{Extended version of Fig. \ref{f:2l}(a) and (b). State $S=1$ is shown in (a), $S=2$ in (b). Intensities are on the left. The second and third columns show top views of the phase-intensity plots, emphasizing the parity with respect to inversion symmetry of wavefunctions $(\psi_a + \psi_b)/\sqrt{2}$ and $(\psi_a - \psi_b)/\sqrt{2}$ in the degenerate subspace.}\label{f:2l_deg}
\end{figure}

In the case of doubly-degenerate excitonic states, the intensity reads $|\Psi(\mathbf{r},\mathbf{r}_h)|^2=|\psi_a(\mathbf{r},\mathbf{r}_h)|^2+|\psi_b(\mathbf{r},\mathbf{r}_h)|^2$. 
Here $\mathbf{r}_h$ is the fixed position of the hole, while $\mathbf{r}$ is the position of the electron. 
What is plotted is the sum of the electron distributions of the two degenerate states $\psi_a$ and $\psi_b$, which are in general complex. 
In order to fully represent the phase of the excitonic wavefunctions, and to give information on the full symmetry of the excitons, it is necessary to rotate states $\psi_a$ and $\psi_b$ in the degenerate subspace until they are both real. 
In Fig. \ref{f:2l_deg}, the phase-intensity plots showing the symmetry with respect to inversion of the lowest-bound Davydov pair in bilayer hBN are shown (cfr. with Fig. \ref{f:2l}(a) and (b)). 
The phase is plotted for states $(\psi_a \pm \psi_b)/\sqrt{2}$ in regions of space where their intensity is relevant (i.e. $|(\psi_a \pm \psi_b)/\sqrt{2}|^2$ is higher than $5\%$ of its maximum value). 
With this representation, both of the degenerate wavefunctions behave in the same way with respect to inversion symmetry as the full exciton.

\FloatBarrier

\bibliography{davydov}

\end{document}